\documentstyle[prl,aps,graphicx]{revtex}

\newcommand{{ \vn }}{\vec{n}}

\newcommand{\beq}{\begin{equation}}
\newcommand{\eneq}{\end{equation}}

\newcommand{\met }{\frac{1}{2}}

\begin{document}

\begin{center}
{\bf {\large Linear Kondo conductance in a quantum dot}}

\vspace*{0.5cm} {\sl Domenico Giuliano}$^{\ddagger }$, {\sl Adele Naddeo$%
^{\dagger \ast }$ and Arturo Tagliacozzo$^{\dagger \ast }$}

\bigskip {\small $^{\ddagger }$ Dipartimento di Fisica, Universit\`{a} della
Calabria, Arcavacata di Rende - Cosenza, Italy}

{\small $~^{\dagger }$ {\sl Coherentia - INFM} (Istituto Nazionale di Fisica
della Materia ), Unit\`{a} di Napoli, Napoli, Italy\\[0pt]
} {\small $^{\ast }$ Dipartimento di Scienze Fisiche Universit\`{a} di
Napoli ''Federico II '',}

{\small Monte S.Angelo - via Cintia, I-80126 Napoli, Italy}

\bigskip
\end{center}

\begin{quotation}
In a tunneling experiment across a quantum dot it is possible to change the
coupling between the dot and the contacts at will, by properly tuning the
transparency of the barriers and the temperature. Gate voltages allow for
changes of the relative position of the dot addition energies and the Fermi
level of the leads. Here we discuss the two limiting cases: weak and strong
coupling in the tunneling Hamiltonian. In the latter case Kondo resonant
conductance can emerge at low temperature in a Coulomb blockade valley. We
give a pedagogical approach to the single channel Kondo physics at
equilibrium and review the Nozi\`{e}res scattering picture of the correlated
fixed point. We emphasize the effect of an applied magnetic field and show
how an orbital Kondo effect can take place in vertical quantum dots tuned
both to an even and an odd number of electrons at a level crossing. We
extend the approach to the two channel overscreened Kondo case and discuss
recent proposals for detecting the non Fermi Liquid fixed point which could
be reached at strong coupling.
\end{quotation}

\bigskip 

\vspace*{0.3cm}

\section{Introduction}

Recently systems have been fabricated which can sustain quantum coherence
because of the smallness of their size, provided that the temperature is low
enough. These mesoscopic systems are nanostructured devices in which quantum
coherence sets in at very low temperatures and modifies the properties of
the device as a whole. This happens notwithstanding the fact that the system
is connected to an external biasing circuit. They have global, geometry
dependent properties and striking quantization phenomena arise which are
largely independent of the specific sample involved: charge quantization (in
unit of the electron charge $e$), periodicity in the magnetic flux quantum $%
\phi _{o}=hc/e$, conductance quantization (in units of $G_{K}=2e^{2}/\pi
\hbar =(6.5K\Omega )^{-1}$).

Among these macroscopic quantum phenomena there is the unitary limit of the
Kondo conductance in tunneling across a quantum dot (QD) at Coulomb Blockade
(CB) \cite{raikh}\cite{aleiner} that was first measured in 1998 \cite
{goldhaber}. The Kondo phenomenon is well known since the sixties and
explains an anomaly in the temperature dependence of the resistivity of
diluted magnetic alloys\cite{kondo}\cite{hewson}.

This review is devoted to some features of equilibrium Kondo conductance
across a quantum dot in the CB regime interacting with two contacts. It is
remarkable here that the dot acts as a single impurity probed by the metal
contacts, so that the properties of the Kondo state are not mediated over
many impurities per cubic centimeter as it happens in diluted magnetic
alloys. A strongly coupled state sets in between the dot and the contacts
and phase coherence is established among the conduction electrons and the
whole structure.

The temperature scale for the interactions between a magnetic impurity and
the delocalized conduction electrons of the host metal is the so called
Kondo temperature $T_{K}$. It is defined as the temperature at which the
perturbative analysis breaks down. Therefore, different approaches are
required to investigate the thermodynamics and the transport properties of a
quantum dot in the Kondo regime in the whole range of temperatures from the
perturbative region down to the unitarity limit.

Recently, the Kondo model and the Anderson impurity model in its Kondo limit
have been deeply investigated by numerical renormalization group (NRG)
calculations \cite{nrg}, the Bethe ansatz method \cite{tsvi} and conformal
field theory (CFT) techniques \cite{ludaff}. Further developments in the NRG
methods have been applied successfully in the crossover region $T\approx
T_{K}$ \cite{bickers}. The zero field spectral and transport properties of
the Anderson model in the Kondo regime \cite{costi} as well as the field and
temperature dependence of the Kondo resonance and the equilibrium
magnetoconductance of the dot \cite{costi1} have been investigated. The
tunneling conductance as a function of the gate voltage has also been
calculated with NRG, in wide temperature range for a single quantum dot with
Coulomb interactions, assuming that two orbitals were active for the
tunneling process \cite{sakai}.

We leave out the case when the electron distribution is not in local
equilibrium about the Kondo impurity and the linear response theory is no
longer sufficient. A number of techniques have been applied to describe
nonequilibrium properties such as the nonlinear conductance, the
nonequilibrium stationary state or the full time development of an initially
out-of-equilibrium system: variational calculations \cite{ng}, perturbation
theory \cite{davies}, equation of motion \cite{meir}, perturbative
functional integral methods \cite{koenig}, non-crossing approximation (NCA)
\cite{meir1}\cite{kroha}, perturbative renormalization group (RG) method in
real time \cite{schoeller}. In particular, the last technique is well suited
to describe quantum fluctuations which are induced by strong coupling
between a small quantum system and an environment. It succeeds in
reproducing the anomalous line shapes of the conductance observed in several
recent experiments \cite{goldhaber}, due to the renormalization of the
resistance and the local energy excitations \cite{schoeller1}.

We also leave out situations in which the levels localized at the dot are
close in energy to the Fermi energy of the contacts (mixed valence models).

The plan of the paper is the following. We start from the tunneling
hamiltonian formalism when the coupling between the dot and the contacts is
weak (Section $II$). A scattering approach in one-dimension is particularly
suitable when studying the linear conductance in the device. The prototype
model to account for strong electron-electron repulsion on the dot is the
Anderson Hamiltonian with onsite repulsion (Section $III$). In the limit of
strong correlation between the dot and the leads the latter model maps onto
the so called ``Kondo Hamiltonian model`` (Section $IV$). A {\sl poor man's
scaling }approach up to second order leads us to the definition of the Kondo
temperature $T_{K}$. Next, we introduce the physics of the single channel
(Section $VI$) and the two channel Kondo problem (Section $VII$), both in
the perturbative region and in the unitarity limit. In such a context we
briefly sketch the Anderson, Yuval and Hamann Coulomb Gas approach \cite
{yuval} to the isotropic and anisotropic one channel Kondo system which
gives a straightforward, though qualitative, insight of the strongly
correlated state.

In the conventional setting the dynamical variable which is coupled to the
electrons propagating from the leads is the total dot spin. There are cases
in which the spin is locked to orbital degrees of freedom or even absent
(``orbital Kondo'') \cite{glazman2}; such cases are better found in a
vertical geometry in presence of a magnetic field orthogonal to the dot \cite
{noi}. A rather strong vertical magnetic field can induce level crossings in
the dot due to orbital effects and produce accidental degeneracies which can
give rise to exotic Kondo couplings (see Section $VI.B-C$). Some attention
is drawn to the cylindrical geometry and to symmetry selection rules in the
cotunneling process, also in connection with proposals for achieving the two
channel Kondo non Fermi liquid fixed point in a vertical dot device \cite
{arturo3}\cite{oreg}\cite{glazman3} (Section $VII.C$).


\section{Tunneling hamiltonian at weak coupling}

Quantum dots (QD) are fabricated in semiconductor heterostructures, by
applying metallic gates to confine few electrons \cite{kouwenhoven}. Because
the confining area is quite small (the radius is $\sim 100\div 1000{{%
\mathring{A}}}$), the charging energy is much larger than the energy
associated with thermal fluctuations, provided temperature is below $%
1^{\circ }K$. Therefore the dot is only weakly linked to metal contacts and
one can bias the system in such a way that the electron number $N$ in the
dot can be changed at will. Because of the confining potential, dots display
a level structure organized in shells, exactly the same as atoms do.
Analyzing this level structure is of primary interest because these few
interacting electrons ($N\leq 20$) confined in a disk-like box (see Fig. 1),
at special values of the parameters, are ruled by strong many-body Coulomb
correlations. Hund's rules can be seen at work: the total spin $S$ of the
electrons confined in the dot is maximized as long as no magnetic field is
applied. On the other hand, a magnetic field in the direction orthogonal to
the disk produces strong orbital effects which favour larger values of the
total angular momentum $M$ as well as total spin $S$.

Quantum dots are remarkable because of Coulomb oscillations. At very low
temperature linear conductance (at vanishing bias $V_{sd}$) is zero, except
for peaks at discrete values of the gate voltage $V_{g}$, when it is
energetically favourable to add one extra electron to the dot. Therefore $%
V_{g}$ controls the number of particles on the dot in the CB regime.

We tune the chemical potential $\mu $ of the left (L) and right (R) bulk
contacts within the CB valley of the conductance at $N$ particles, that is $%
\mu _{N}<\mu <\mu _{N+1,\alpha }\equiv ^{N+1}\!E_{\alpha }-^{N}\!E_{0}$;
here the energies $^{N\pm 1}\!E_{\alpha }$ are the total energies for the
dot with $N\pm 1$ particles characterized by the quantum numbers $\alpha $
and $^{N}\!E_{0}$ is the ground state energy (GS) at $N$ particles. If $N$
is odd, then the GS is at least doubly degenerate because the odd particle
state can have spin $\sigma =\uparrow ,\downarrow $. In this case the dot
can be treated as a single Anderson impurity.

The peculiarity of these artificial atoms is in that the level structure can
be investigated by measuring the tunneling current. Peaks in the linear
conductance versus gate voltage $V_{g}$ separate regions at Coulomb blockade
(CB) in which $N$ is fixed and differs by just one electron. The peaks at
zero source-drain voltage occur whenever $V_{g}$ compensates the required
chemical potential for electron addition or subtraction. Hence, measuring a
tunnel current across the device $vs$ $\ V_{g}$ provides the spectroscopy of
the dot levels. Adding a magnetic field the spin state of the dot can be
changed, which in turn changes selection rules for electron tunneling \cite
{arturo1}.

The height and the width of the conductance peaks at resonance depends on
temperature $T$. Let the dimensionless conductance be $g$ (in units of $2%
\frac{e^{2}}{h}$ (factor of $2$ is for the spin)). In the simplest case the
maximum of the conductance at the resonant peak is $g_{max}\propto \Gamma
/(k_{B}T)$ and the halfwidth of the peak $\Gamma =\pi \nu \left( 0\right)
\left| V\right| ^{2}$ is $\propto T$; here $V$ is the tunneling strength
(see below) and $\nu (0)$ is the density of states at the Fermi energy. In
the CB region tunneling via virtual states of the dot with $N\pm 1$ is only
fourth order perturbation in $V$ (they are named cotunneling processes), $%
g\propto e^{2}\pi \left( \frac{2\nu \left( 0\right) |V|^{2}}{\epsilon _{d}}%
\right) ^{2}/\hbar $, and vanishes exponentially with reducing temperature.

Nevertheless, unexpectedly, transport measurement in a dot gives rise to
some new physics which can be related to the Kondo effect of spin impurities
in non magnetic metal alloys (see Fig. 2) \cite{goldhaber}.

In the following we introduce the model for a quantum dot interacting with
the contacts and define an equivalent one-dimensional Hamiltonian.

\subsection{Current within the tunneling Hamiltonian}

In this Subsection we discuss the mutual influence between the contacts and
the dot, starting from the weak link limit. Conduction electrons in the
leads constitute a Fermi sea (FS) of non interacting fermions with plane
wave single particle wavefunctions on side $L$ and on side $R$. The dot is
described by an Hamiltonian $H_{D}$ and coupling between the dot and the
leads is accounted for through a tunneling term, so that the Hamiltonian
describing the whole system is:

\begin{equation}
H=H_{D}+\sum_{k\sigma }\epsilon _{k\sigma }^{R}a_{k\sigma }^{\dagger
}a_{k\sigma }+\sum_{k\sigma }\epsilon _{k\sigma }^{L}b_{k\sigma }^{\dagger
}b_{k\sigma }+\sum_{k\alpha \sigma }[V_{k\alpha }c_{k\alpha \sigma
}^{\dagger }d_{\sigma }+V_{k\alpha }^{\ast }d_{\sigma }^{\dagger }c_{k\alpha
\sigma }].  \label{ap1}
\end{equation}
The right and left Fermi sea (FS) of the two contacts $R,L$ are acted on by
operators $c_{k R \sigma }\equiv a_{k\sigma }$ and $c_{k L \sigma } \equiv
b_{k\sigma }$ ( here the index $\alpha $ stands for $R,L$). The canonical
transformation of the dot problem made by Glazman and Raikh \cite{raikh} is
just the construction of two species of fermions whose wavefunction is even
or odd with respect to the dot center, in the case that the two barriers are
unequal $V_{\alpha }=V_{R},V_{L}$. It changes the picture from operators $%
a_{k\sigma }$ and $b_{k\sigma }$ to operators $\alpha _{k\sigma }$ and $%
\beta _{k\sigma }$ given by
\begin{eqnarray}
\alpha _{k\sigma } &=&ua_{k\sigma }+vb_{k\sigma },\text{ \ \ }\beta
_{k\sigma }=ub_{k\sigma }-va_{k\sigma }  \nonumber \\
u &=&\frac{V_{R}}{V},\text{ \ }v=\frac{V_{L}}{V},\hspace*{0.4cm}V=\sqrt{%
|V_{R}|^{2}+|V_{L}|^{2}},\hspace*{0.4cm}\Gamma _{R,L}=\pi \left|
V_{R,L}\right| ^{2}\nu (0)  \label{tu1}
\end{eqnarray}
where $\nu (0)$ is the density of states at the Fermi energy.

A general formula for the the conductance of interacting systems \cite{ng1}
\cite{meir}\cite{jauho} has been written, resorting to the nonequilibrium
Keldysh formalism \cite{noneq}; in such a framework the current through the
interacting region is written in terms of the distribution functions in the
leads and local properties of the intermediate region, such as the
occupation and the density of states at the dot site:
\begin{equation}
J=\frac{ie}{2h}\int d\omega \left( Tr\left\{ 2\left[ f_{L}(\omega )\Gamma
_{L}-f_{R}(\omega )\Gamma _{R}\right] \left( {\bf G}^{r}-{\bf G}^{a}\right)
\right\} +Tr\left\{ 2\left( \Gamma _{L}-\Gamma _{R}\right) {\bf G}%
^{<}\right\} \right) ,  \label{currmeir}
\end{equation}
where ${\bf G}^{r}$, ${\bf G}^{a}$, ${\bf G}^{<}$ are the usual retarded,
advanced, Keldysh Green functions for the dot in interaction with the leads
and the $f$ are the Fermi functions. The Green's function ${\bf G}^{r}$ will
be denoted as $G_{dd}$ in the following. Both ${\bf G}^{r}$ and ${\Gamma }%
_{R,L}$ are matrices in case that many channels are present.

Formula (\ref{currmeir}) can be cast in a more simple form in the case that
the couplings to the leads differ only by a constant factor \cite{meir}:
\begin{equation}
J=-2\frac{e}{h}\int d\omega \left[ f_{L}(\omega )-f_{R}(\omega )\right] \Im
mTr\left\{ \tilde{\Gamma}{\bf G}^{r}(\omega )\right\} ,  \label{condumeir}
\end{equation}
where $\tilde{\Gamma}=\Gamma _{R}\Gamma _{L}/\Gamma $. Here $\Gamma = \Gamma
_{R}+ \Gamma _{L}$.

Now we derive explicitely $G_{dd}$ within the lowest order perturbation in
the tunneling.

We start from the Hamiltonian in eq. (\ref{ap1}), where $H_{D}$ is given by
a single impurity energy level $\epsilon _{d}$. All quantities will be
scalar quantities for simplicity. We define the electron retarded Green's
function for the unperturbed leads: $G_{0}^{-1}=\sum_{k}(\omega -\epsilon
_{k}+i0^{+})$. Projecting the equations for the total Green's function:
\begin{eqnarray}
(i\omega _{n}-H)G(i\omega _{n}) &=&{\bf 1}  \nonumber \\
G^{\dagger }(i\omega _{n})(-i\omega _{n}-H) &=&{\bf 1}
\end{eqnarray}
onto states in which one single extra particle is occupying the delocalized
state $|k>$ or the impurity state $|d>$ we have:
\begin{eqnarray}
(i\omega _{n}-\epsilon _{k})G_{k,k^{\prime }\sigma }(i\omega _{n}) &=&\delta
_{kk^{\prime }}+V_{k}G_{d,k^{\prime }\sigma }(i\omega _{n})  \nonumber \\
G_{d,k^{\prime }\sigma }(i\omega _{n})(-i\omega _{n}-\epsilon _{k^{\prime}})
&=&G_{d,d\sigma }(i\omega _{n})V_{k^{\prime }}^{\ast }
\end{eqnarray}
where $G_{d,k\sigma }(i\omega _{n})=<d|G|k>$, $G_{k,k\sigma }(i\omega
_{n})=<k|G|k>$, $G_{d,d\sigma }(i\omega _{n})=<d|G|d>$ and $V_k = \langle k
|V| d \rangle $.

This gives for each scattering channel (here in order to simplify the
notation we suppress the channel label $\alpha $):
\begin{eqnarray}
G_{k,k^{\prime }\sigma }(i\omega _{n}) &=&\frac{\delta _{kk^{\prime }}}{%
i\omega _{n}-\epsilon _{k}}+\frac{V_{k}}{i\omega _{n}-\epsilon _{k}}%
G_{d,d\sigma }(i\omega _{n})\frac{V_{k^{\prime }}^{\ast }}{i\omega
_{n}-\epsilon _{k^{\prime }}}  \nonumber \\
&\equiv &G_{k\sigma }^{0}\delta _{kk^{\prime }}+G_{k\sigma
}^{0}V_{k}G_{d,d\sigma }V_{k^{\prime }}^{\ast }G_{k^{\prime }\sigma ^{\prime
}}^{0}.  \label{greenk}
\end{eqnarray}

The density of states of the scattering electrons is defined as:
\begin{equation}
\nu (\omega )=-\frac{1}{\pi }\Im m\sum_{k}{\bf G}_{k,k\sigma }^{r}(\omega ).
\end{equation}

Using similar equations for the dot, we write:
\begin{equation}
G_{d,d\sigma }(i\omega _{n})=\frac{1}{i\omega _{n}-\epsilon
_{d}-\sum_{k}|V_{k}|^{2}/(i\omega _{n}-\epsilon _{k})}.  \label{greend}
\end{equation}
Now, let us consider the $L$ and $R$ contacts as two equal Fermi gases of
noninteracting particles and equal chemical potential $\mu =0$ (within
linear conductance regime a source drain voltage $V_{sd}$ is not applied).
In such a case it is enough to discuss a single effective contact and the
corresponding wavefunctions are plane waves of wavevector $k$ in the $x$%
-direction with a label $\alpha $ which includes all other quantum numbers.
Their energy dispersion $\epsilon _{k\alpha }$ can be linearized about $\mu $
with $\epsilon _{q}\approx \hbar v_{F}q$, where $q$ is the momentum measured
with respect to the Fermi momentum and $v_{F}$ is the Fermi velocity. Using
a constant density of states $\nu (0)$ (at the Fermi energy per spin $%
L_{o}/2\pi \hbar v_{F}$ where $L_{o}$ is the size of the system) and a
bandwidth of size $2D$ and neglecting the $k$ dependence of $V_{k}$, the sum
in eq. (\ref{greend}) is readily done:
\begin{equation}
\sum_{k}\frac{|V_{k}|^{2}}{i\omega _{n}-\epsilon _{k}}=|V|^{2}\nu
(0)\int_{-D}^{D}d\epsilon \frac{1}{i\omega _{n}-\epsilon }=\frac{\Gamma }{%
\pi }\ln \frac{i\omega _{n}+D}{i\omega _{n}-D}\equiv \Sigma _{d}(i\omega
_{n}).  \label{selfdot}
\end{equation}
The retarded Green's function for the dot is obtained from $G_{d,d\sigma
}(i\omega _{n})$ in the limit to real frequencies: $i\omega _{n}\rightarrow
\omega +i0^{+}$ according to:
\begin{equation}
{\bf G}_{d,d\sigma }^{-1}(i\omega _{n})=\left\{ i\omega _{n}-\epsilon
_{d}-\Sigma _{d}(i\omega _{n})\right\} \rightarrow \omega -\epsilon _{d}-%
\frac{\Gamma }{\pi }\ln \left| \frac{v_{F}D+\omega }{v_{F}D-\omega }\right|
+i\Gamma =\omega -\widetilde{\epsilon }_{d}+i\Gamma ,  \label{greendot}
\end{equation}
\noindent where $\widetilde{\epsilon }_{d}$ is the renormalized dot energy.
The coupling of the dot to the contacts shifts the location of the pole
corresponding to the energy of the localized level and gives a finite width $%
\Gamma $ to the resonance.

\subsection{1-d Scattering formalism}

At zero temperature scattering is only elastic. In the following we develop
a $1-d$ scattering approach which is mostly useful in vertical structures
\cite{erio}, next we show that also the tunneling conduction can be cast
into the scattering formalism.

If the evolution operator $U(t,t^{\prime })$ is known, the transmission
amplitude can be written in a scattering approach as $\theta ^{R\rightarrow
L}=<b^{\dagger }U(+\infty ,-\infty )a>=v^{\ast }u<\alpha ^{\dagger }(\infty
)\;\alpha (-\infty )>-u^{\ast }v<\beta ^{\dagger }(\infty )\;\beta (-\infty
)>\equiv v^{\ast }uS^{0}+u^{\ast }vS^{1}$, where we have defined the two
expectation values as the scattering matrix elements for the two uncoupled
even and odd channels. Hence, the conductance takes the form of the Landauer
formula:
\begin{equation}
g=T^{R\rightarrow L}\equiv \left| \theta ^{R\rightarrow L}\right|
^{2}=4\left| v^{\ast }u\right| ^{2}\left| \met\sum_{l}S^{l}\right| ^{2}=%
\frac{4\Gamma _{R}\Gamma _{L}}{(\Gamma _{R}+\Gamma _{L})^{2}}\left| \met%
\sum_{l}S^{l}\right| ^{2}.  \label{trans}
\end{equation}
The potential after the transformation of eq. (\ref{tu1}) has become even,
hence unitarity is satisfied by each channel individually: $|S^{l}|^{2}=1$, $%
l=0,1$. The unitary limit of the conductance is $g_{u}=\frac{4\Gamma
_{R}\Gamma _{L}}{(\Gamma _{R}+\Gamma _{L})^{2}}$. In particular, if the
potential is $\delta (x)$, odd parity is totally transmitted and such that $%
S^{1}=-1$.

Eq. (\ref{trans}) is valid for a general system with an interacting
intermediate region. One can match this approach to a 1-d scattering
approach for non interacting electrons as it is done here below.

Let us place the impurity (QD) at $x=0$ and consider the scattering
amplitude $f_{L,R}$ of a one electron wavefunction $\psi (x)$
\begin{eqnarray}
\psi _{>}(x) &\propto &e^{ikx}+f_{R}e^{ikx}\hspace*{0.5cm}x>>0  \nonumber \\
\psi _{<}(x) &\propto &e^{ikx}+f_{L}e^{-ikx}\hspace*{0.5cm}x<<0.
\end{eqnarray}

Here the transmission coefficient is $T=|1+f_{R}|^{2}$ while the reflection
coefficient is $R=|f_{L}|^{2}$ and they satisfy the conservation of flux: $%
T+R=1$.

If the dot structure is even with respect to the origin along the vertical
axis, the even parity $l=0$ and the odd parity $l=1$ channels are
independent. It is then useful to define even and odd parities $f^{l}$: $%
f^{0}=\met(f_{L}+f_{R})$, $f^{1}=\met(f_{R}-f_{L})$ and the elastic $t-$
matrix $t^{l}=ik_{o}f^{l}/\pi $. Here the energy of the incoming particle is
$\hbar v_{F}k_{o}$ (in units $\hbar v_{F}=1$, it follows that $[t]=energy$)
and $k_{o}=2\pi /L_{o}$ (where $L_{o}$ is the linear size of the system).
The $t-$ matrix is related to the $S-$ matrix according to:
\begin{eqnarray}
S^{l} &=&e^{2i\delta ^{l}};\hspace*{0.5cm}S^{l}-1=-\frac{2\pi i}{k_{o}}t^{l}
\nonumber \\
t^{l} &=&-\frac{k_{o}}{\pi }\sin \delta ^{l}e^{i\delta ^{l}}
\end{eqnarray}
where $\delta ^{l}$ are the phase shifts for the two parities. In this
context unitarity of the $S-$matrix, $\met\sum_{l}|S^{l}|^{2}=1$, is the
same as flux conservation $R+T=1$.

Conductance is given by the Landauer formula:
\begin{equation}
conductance=g=T=\left| 1+f_{R}\right| ^{2}=\left| 1+i\sum_{l}\sin \delta
^{l}e^{i\delta ^{l}}\right| ^{2}.  \label{condu}
\end{equation}
In the case of resonant tunneling we have $T\sim 1$ and $R=|f_{L}|^{2}\sim 0$%
, so that eq. (\ref{condu}) becomes $g\rightarrow 1$, that is the unitary
limit.

The condition $R=|\pi (t^{0}-t^{1})/(ik_{o})|^{2}=\sin ^{2}(\delta
^{0}-\delta ^{1})\sim 0$ yields $\delta ^{0}\sim \delta ^{1}\equiv \delta
mod\;\pi $, while in eq. (\ref{condu}) we have $T\sim 1$ for $\delta
^{0}=\delta ^{1}\equiv \delta \rightarrow \pi /2$.

If the potential is even ($\Gamma _{R}=\Gamma _{L}$), unitarity is satisfied
by each channel individually: $|S^{l}|^{2}=1$, $l=0,1$; in particular, if
the potential is $\delta (x)$, odd parity is totally transmitted and such
that $S^{1}=-1$, and the conductance becomes:
\begin{equation}
g=\left| \met(S^{0}-1)\right| ^{2}=\left| \frac{\pi }{k_{o}}t^{0}\right|
^{2}=sin^{2}\delta ^{0}.  \label{gdot}
\end{equation}
Again, if $\delta ^{0}=\pi /2$ conductance reaches the unitarity limit.

In the following we describe the basic approximations and the weak coupling
limit of very opaque barriers between dot and contacts.

In this case tunneling of lead electrons across the dot is a perturbative
process as derived in Section IIA. Within such a perturbative regime neither
the dot nor the contacts are much affected by the interaction with the
other. The relevant effect on the QD is the shifting of its levels and a
level broadening, which is a second-order effect in the tunneling strength $V
$. This shows itself as a small imaginary part added to the energy levels,
i.e. a finite lifetime.

It is easy to show that the linear conductance given above can be rephrased
in terms of the imaginary part of the Green's function as in eq. (\ref
{condumeir}) at the perturbative level. To this end, let us first state some
formalities regarding the selfenergy ${\bf \Sigma }$ and the ${\bf t}$
matrix:

$(E-H_{0}-\Sigma ){\bf G}={\bf 1}$; ${\bf G}_{0}(E-H_{0})={\bf 1}\rightarrow
{\bf G}_{0}{\bf G}^{-1}+{\bf G}_{0}{\bf \Sigma }={\bf 1}\rightarrow {\bf G}=%
{\bf G}_{0}+{\bf G}_{0}{\bf \Sigma G}$.

In the Born approximation $\Sigma $ and ${\bf t}$ coincide because by
definition is ${\bf G}={\bf G}_{0}+{\bf G}_{0}{\bf t}{\bf G}_{0}$. We assume
that in tunneling the odd parity channel is fully transmitted ($S^{1}=-1$)
and the even parity one gives (see eq. (\ref{greenk})):
\begin{eqnarray}
t^{0} &\sim &VG_{d,d}V,\hspace*{0.5cm}G_{d,d}=\frac{1}{\epsilon -\widetilde{%
\epsilon }_{0}+i\Gamma }  \nonumber \\
&\rightarrow &t^{0}\sim \frac{|V|^{2}}{\epsilon -\widetilde{\epsilon }%
_{0}+i\Gamma },\hspace*{0.5cm}\frac{\Im mt^{0}}{\Re et^{0}}=\tan \delta ^{0}=%
\frac{-\Gamma }{\epsilon -\widetilde{\epsilon }_{0}},  \nonumber \\
\sin ^{2}\delta ^{0} &=&\frac{\tan ^{2}\delta ^{0}}{1+\tan ^{2}\delta ^{0}}=%
\frac{\Gamma ^{2}}{(\epsilon -\widetilde{\epsilon }_{0})^{2}+\Gamma ^{2}}.
\label{born}
\end{eqnarray}
Here $V$ is the tunneling matrix element defined in eq. (\ref{tu1}), $%
\widetilde{\epsilon }_{0}$ is the renormalized quantity defined in eq. (\ref
{greendot}) and $\Gamma =\Gamma _{R}+\Gamma _{L}$ is the inverse lifetime of
the resonance. This implies that eqs. (\ref{trans},\ref{gdot}) become:
\begin{equation}
g={g}_{u}\sin ^{2}\delta ^{0}=\frac{4\Gamma _{R}\Gamma _{L}}{(\epsilon -%
\widetilde{\epsilon }_{0})^{2}+\Gamma ^{2}},  \label{condu1}
\end{equation}
where $g_{u}=\frac{4\Gamma _{R}\Gamma _{L}}{(\Gamma _{R}+\Gamma _{L})^{2}}$.
On the other hand, because $\Im mG=-\Gamma /[(\epsilon -\widetilde{\epsilon }%
_{0})^{2}+\Gamma ^{2}]$, eq. (\ref{condumeir}) becomes:
\begin{equation}
g=\int d\omega \left( -\frac{\partial f(\omega )}{\partial \omega }\right)
\tilde{\Gamma}\frac{\Gamma }{(\epsilon -\widetilde{\epsilon }%
_{0})^{2}+\Gamma ^{2}},  \label{acondu1}
\end{equation}
where $\tilde{\Gamma}=\Gamma _{R}\Gamma _{L}/\Gamma $. At zero temperature
both results of eqs. (\ref{condu1}) and (\ref{acondu1}) coincide.

At finite temperature, if the odd channel is fully transmitted eq.(\ref
{trans}) can be generalized as
\begin{equation}
g={g}_{u}\int d\omega \left( -\frac{\partial f(\omega )}{\partial \omega }%
\right) \left[ -\pi \nu (0)\Im m\left( t^{0}\right) \right]  \label{correc1}
\end{equation}
where $t^{0}$ is the ${\bf t}$ matrix above defined and related to the exact
retarded Green function through the relation ${\bf G}={\bf G}_{0}+{\bf G}_{0}%
{\bf t}{\bf G}_{0}$. We have used the optical theorem:
\begin{equation}
\frac{\pi ^{2}}{k_{o}^{2}}Tr\left\{ t^{0}t^{0\dagger }\right\} =\Re e\left\{
\frac{i\pi }{k_{o}}t^{0}\right\}  \label{optical}
\end{equation}
which follows from the unitarity condition $|S^{0}|^{2}=1$.

\section{Role of the onsite repulsion $U$ in tunneling}

Up to now $H_{D}$ referred just to a single impurity level $\epsilon _{d}$.
Indeed, charging energy $U$ is the main feature of a $QD$ and we have to
deal with it. Something which is closer to a $QD$ is an impurity level with
onsite repulsion $U$. Inclusion of $U$ in the Hamiltonian (\ref{ap1}) leads
to the single level Anderson model:

\begin{equation}
H_{{\rm AND}}=H_{{\rm lead}}+\epsilon _{d}\sum_{\sigma }n_{\sigma
}+U\sum_{\sigma \sigma ^{^{\prime }}}n_{\sigma }n_{\sigma ^{^{\prime
}}}+\sum_{k\alpha \sigma }V_{\alpha }[c_{k\alpha \sigma }^{\dagger
}d_{\sigma }+d_{\sigma }^{\dagger }c_{k\alpha \sigma }].  \label{app1}
\end{equation}

If $\epsilon _{d}=-U/2$ with respect to the chemical potential of the
conduction electrons $\mu $ (taken as the zero of the single particle
energies), the Anderson model which arises is symmetric. In fact, the
energies of the empty impurity state,$^{0}E$, and of the doubly occupied
impurity state, $^{2}E=2\epsilon _{d}+U$, are both zero, while the singly
occupied impurity level has energy $^{1}E=-U/2$.

In order to understand how the Coulomb repulsion on the dot site affects the
Green's function of the dot we use a path integral formalism. We show that,
in the limit in which the charge degree of freedom on the dot is frozen ($%
U\rightarrow \infty $), the dot Green's function describes the dynamics of
the only degree of freedom left, that is the dot magnetization $<n_{\uparrow
}-1/2>\equiv <1/2-n_{\downarrow }>$ which is forced by a stochastic field $%
X(\tau )$ in imaginary time, produced by the scattering of the lead
electrons, with a gaussian probability distribution. The excitation energy
associated to it is shifted from $\epsilon _{d}$ to the Fermi level: this is
the origin of the resonance at the Fermi level in the Kondo problem.

As a first step, we will rephrase the previous result of an impurity with $%
U=0$ in this new approach. After linearizing the bands, the leads action can
be written in terms of the field operators $\psi _{u\alpha }$ (where $\alpha
=L,R$ and $u$ includes all other quantum numbers) as:
\begin{equation}
S_{{\rm lead}}=-iv_{F}\int_{0}^{\beta }d\tau \int dx\sum_{u}\sum_{\alpha
=L,R}\;\psi _{u\alpha }^{\dagger }(x,\tau )\frac{d}{dx}\psi _{u\alpha
}(x,\tau ).  \label{reasl}
\end{equation}
\noindent

With respect to the tunneling action, if the $L$ and $R$ barriers are equal
and the dot is zero dimensional the interaction term only includes the
symmetric combinations $\Phi _{u}(\tau )=\frac{1}{\sqrt{2}}(\psi
_{uL}(x=0,\tau )+\psi _{uR}(x=0,\tau ))$ at the origin:
\begin{equation}
S_{{\rm T}}=\int_{0}^{\beta }d\tau \sqrt{2}\sum_{u}[V\Phi _{u}^{\dagger
}(\tau )d_{u}(\tau )+V^{\ast }d_{u}^{\dagger }(\tau )\Phi _{u}(\tau )].
\label{newint}
\end{equation}

The total action is:
\begin{equation}
{\cal {A}}=\int_{0}^{\beta }d\tau \left\{ \sum_{u}d_{u}^{\dagger }(\tau )%
\left[ \frac{\partial }{\partial \tau }+(\epsilon _{d}-\mu )\right]
d_{u}(\tau )\right\} +S_{{\rm lead}}+S_{{\rm T}}.
\end{equation}
One can first integrate out the degrees of freedom of the $\psi _{u\alpha
}(\tau ,x)$ fields for $x\neq 0$ which are free-like, and next the field $%
\Phi _{u}$ at $x=0$ which is the only one interacting with the dot variable $%
d_{u}(\tau )$. Because the fields in the leads are non interacting, the
result of the gaussian integration yields an effective action for the dot:

\begin{equation}
-S_{{\rm D}}^{{\rm Eff}}\propto \ln \left\{ \int \prod_{\alpha =L,R}\prod_{u}%
{\cal D}\psi _{u\alpha }\;{\cal D}\psi _{u\alpha }^{\dagger }\;e^{-{\cal {A}}%
}\right\} =-\beta \sum_{i\omega _{n}}d_{u}^{\dagger }(i\omega _{n})\left(
i\omega _{n}-\epsilon _{d}-\Sigma (i\omega _{n})\right) d_{u}(i\omega _{n})
\label{sef}
\end{equation}
\noindent where the self-energy correction to the Green's function of the
dot $\Sigma (i\omega _{n})$ = $\frac{\Gamma }{\pi }\ln (\frac{i\omega _{n}+D%
}{i\omega _{n}-D})$ was obtained in eq. (\ref{selfdot}).

\subsection{Coulomb repulsion on the dot: freezing of the charge degree of
freedom}

We now discuss the role of the onsite Coulomb interaction. In the large-$U$
limit we have:
\begin{eqnarray}
\exp \int_{0}^{\beta }d\tau \left\{ -\epsilon _{d}(n_{\uparrow
}+n_{\downarrow })-Un_{\uparrow }n_{\downarrow }\right\} &=&e^{-\frac{U}{4}%
\int_{0}^{\beta }d\tau \left\{ (n_{\uparrow }+n_{\downarrow })^{2}+\frac{4}{U%
}\epsilon _{d}(n_{\uparrow }+n_{\downarrow })\right\} }{{\cdot} }e^{\frac{U}{4}%
\int d\tau (n_{\uparrow }-n_{\downarrow })^{2}}  \nonumber \\
&=&e^{\frac{\beta }{U}\epsilon _{d}^{2}}\;\delta (n_{\uparrow
}+n_{\downarrow }+2\epsilon _{d}/U){{\cdot} }e^{\frac{U}{4}\int_{0}^{\beta
}\;d\tau (n_{\uparrow }-n_{\downarrow })^{2}},  \label{hub}
\end{eqnarray}
where the delta function, which is defined by the last equality in the limit
$U\rightarrow \infty $, implements the constraint of single site occupancy
in the symmetric case, $\epsilon _{d}=-U/2$.

The quartic interaction is decoupled by means of a Hubbard-Stratonovitch
boson field $X ( \tau )$, according to the identity:

\[
e^{\frac{U}{4}\int_{0}^{\beta }d\tau (n_{\uparrow }-n_{\downarrow
})^{2}}=\int DXe^{-\frac{1}{4U}\int_{0}^{\beta }d\tau (X^{2}(\tau
)+2U(n_{\uparrow }-n_{\downarrow })X(\tau ))}.
\]
Having introduced the field $X(\tau )$, the partition function ${\cal Z}(\mu
)$ takes the form:

\begin{eqnarray}
{\cal {Z}}(\mu ) &\propto &\int DXe^{-\frac{1}{4U}\int_{0}^{\beta }d\tau
X^{2}(\tau )}  \nonumber \\
&&{{\times} }\int \Pi _{i}\left( Dd_{i}Dd_{i}^{\dagger }e^{-\int_{0}^{\beta
}d\tau d\tau ^{\prime }d_{i}^{\dagger }(\tau )G_{(0)}^{-1}(\tau -\tau
^{\prime })d_{i}(\tau ^{\prime })}\right)  \nonumber \\
&&{{\times} }\delta (n_{\uparrow }+n_{\downarrow }-1){{\cdot} }\met%
\sum_{j=\uparrow ,\downarrow }e^{(-1)^{j}\int_{0}^{\beta }d\tau \lbrack
n_{j}-\met]X(\tau )}.  \label{part2}
\end{eqnarray}
Note that now the term $\epsilon _{d}\sum_{i}n_{i}$ was included in eq. (\ref
{hub}), so in this case the dot Green's function is $G_{(0)}^{-1}(i\omega
_{n})=i\omega _{n}-\Sigma _{(0)}(i\omega _{n})$. At odds with the case $U=0$
here the resonance is at the Fermi level, in spite of the fact that the
original localized level is at $\epsilon _{d}$. The partition function in
eq. (\ref{part2}) describes an effective spin-1/2 coupled to the fluctuating
magnetic field $X(\tau )$; its dynamics is constrained by the requirement
that the impurity is singly occupied. It could be shown that the single site
occupancy constraint is fulfilled in the average when the partition function
of eq. (\ref{part2}) is used.

\subsection{Quenching of the magnetic moment: singlet ground state}

The representation of the partition function given in eq. (\ref{part2})
allows us to recognize the doubly degenerate ground state (GS) of the
impurity spin $S_{d}=1/2$ with $S_{d}^{z}=\left( -1\right) ^{\sigma
}(n_{\sigma }-1/2)$, $\sigma =\uparrow ,\downarrow $, driven by the field $%
X(\tau )$ and produced by virtual tunneling of electrons on and off the dot
at energy $\mu $.

Anderson, Yuval and Hamann \cite{yuval} calculated eq. (\ref{part2}) by
integrating out the impurity ($d,d^{\dagger }$ fields) and showing that the
interaction with the delocalized electrons gives rise to a dynamics of the
field $X(\tau )$ which also interacts with itself at different times
according to a logarithmic law. The problem was solved by mapping eq. (\ref
{part2}) onto the partition function of a Coulomb gas (CG) \cite{ni} of
flips in $1-d$, as we show in some detail in this subsection.

Let us define the new field $\xi (\tau )=X(\tau )/\Gamma $, where $\Gamma
=\pi \nu (0)\left| V\right| ^{2}$ is related to the tunneling strength. At
low temperatures, saddle point solutions of the resulting single particle
effective action are sequences of instantons $\xi _{\pm }(\tau ,l)=\pm \xi
_{M}\tanh ((\tau -\tau _{l})/\tau _{M})$ (where $\tau _{l},l=1,2,\ldots $
are the centers and $\tau _{M}\sim \frac{1}{\epsilon _{F}}$ is the width,
that is some cut-off which regularizes the theory at short times)
corresponding to jumps between the two minima of the effective potential $%
V_{eff}[\xi ]=\left( \Gamma ^{2}/U\right) \xi ^{2}-2\Gamma /\pi \left[ \xi
{\rm \tan }^{-1}(\xi )-\frac{1}{2}\ln (1+\xi ^{2})\right] $, located at $\xi
_{M}=\pm U/2\Gamma $, and interacting via a logarithmic potential $\alpha
^{2}\ln |(\tau _{i}-\tau _{j})/\tau _{M}|$.

For such a CG we can define the bare strength $\alpha _{b}^{2}$ of the
logarithmic interaction (the ''charge'') as $\alpha _{b}^{2}=2(1-8\Gamma
/U\pi )$ and the ``fugacity'' $Y$ as $Y\equiv \tau _{M}e^{-S_{{\rm tot}}}$.

Thus, the full partition function may be approximated with the sum over the
trajectories given by hopping paths and will be written as:

\begin{equation}
Z=\sum_{N=0}^{\infty }\frac{1}{\left( 2N\right) !}\int_{0}^{\beta }\frac{%
d\tau _{2N}}{\tau _{M}}\ldots \int_{0}^{\tau _{2}-\tau _{M}}\frac{d\tau _{1}%
}{\tau _{M}}[e^{\frac{1}{2}\sum_{i\neq j=1}^{2N}(-1)^{i+j}\alpha ^{2}\ln
\left| \frac{\tau _{i}-\tau _{j}}{\tau _{M}}\right| }Y^{2N}],  \label{part1}
\end{equation}
that is the partition function of an effective one-dimensional gas of spin
flips. The integral over the ``centers of the instantons'' has to be
understood such that $\tau _{i}$ and $\tau _{j}$ never become closer than $%
\tau _{M}$.

Now, we are ready to perform the RG analysis to get the behavior of the
model at large time scales (low temperatures) \cite{yuval}. The bare
fugacity of the CG is $Y_{b}=\tau _{M}\exp (-\bar{{\cal {A}}})$ where $\bar{%
{\cal {A}}}\sim \tau _{M}U$ is the action of one single flip. The scaling of
the fugacity and the renormalization of the coupling constant induced by
processes of fusion of charges lead to the following Renormalization Group
(RG) equations \cite{ni}:

\begin{equation}
\frac{dY}{d\ln \tau _{M}}=(1-\frac{\alpha ^{2}}{2})Y\text{ };\frac{d\alpha
^{2}}{d\ln \tau _{M}}=-2Y^{2}\alpha ^{2}  \label{rg1}
\end{equation}
(see Appendix A for a sketch on the derivation).

We see clearly that the flow is towards $Y\rightarrow \infty $ and $\alpha
^{2}\rightarrow 0$ and the scaling invariant energy (that is the Kondo
temperature which we will introduce in the next section) is $T_{K}=\tau
_{M}^{-1}e^{-1/(1-\frac{\alpha ^{2}}{2})}\sim (U\Gamma )^{\met}e^{-\pi
U/(8\Gamma )}$.

Condensation of instantons in the doubly degenerate GS leads to the Kondo
singlet $<S^{z}>=0$ on the dot. Now we present an heuristic argument for
such quenching of the total spin $S$ on the dot; it runs as follows.

The dynamics of the field $\xi (\tau )$ with action $\bar{{\cal {A}}}$ can
be mimicked by a two-level system hamiltonian $H_{2l}$ with hopping energy $%
\lambda \sim \met m_{eff}\left( \frac{d\xi }{d\tau }\right) ^{2}\sim \met
\frac{\Gamma ^{2}}{U}\tau _{M}^{2}(\xi _{M}/\tau _{M})^{2}\sim U/2$. The
role of the interaction is to project out higher energy components from the
dot states. Let us denote by $|\pm >$ the two eigenstates of $H_{2l}$, i.e.
the singlet and the triplet (with respect to the total spin of the dot plus
conduction electrons, $S=0$ and $S=1$) on the dot. Instantons, by flipping
the impurity spin, produce a dynamics of the system ``dot + conduction
electrons `` between these two states.

Now, let us make an interesting analogy with thermodynamics. At temperature $%
\beta ^{-1}$ the probabilities of having the system in one of the two states
will be given by \cite{arturo2}:

\[
P_{+}=\frac{1}{1+e^{-2\beta \lambda }}\;\;;\;P_{-}=\frac{e^{-2\beta \lambda }%
}{1+e^{-2\beta \lambda }}
\]
partly eliminating the component on the high energy eigenstate $|->$. Thus
we are ready to give the connection with the CG picture described by the
partition function of eq. (\ref{part1}). In our case the dynamics is given
by the coherent zero point fluctuations of the system as a whole. From the
statistical weight associated to a configuration with $N$ instantons we can
easily write down the formula:
\begin{equation}
\langle N\rangle \equiv \langle N_{{\rm inst}}\rangle =\frac{%
\sum_{N=0}^{\infty }2NY^{2N}/(2N)!}{\sum {}_{N=0}^{\infty }Y^{2N}/(2N)!}=Y%
\frac{d}{dY}\ln ({\cosh }Y)=Y{\tanh }Y.  \label{ave2}
\end{equation}
The frequency $\lambda /2\pi $ is obviously related to the average number of
flips: $\beta \lambda /2\pi =<N>=Y\tanh Y$, as a direct implication of eq. (%
\ref{ave2}).

Hence, the probabilities of having the system in the states $|\pm >$ within
the ground state ($T=0$) are:
\begin{equation}
P_{+}=\frac{1}{1+e^{-4\pi Y\tanh Y}}\:\:;\:\:P_{-}=\frac{e^{-4\pi Y\tanh Y}}{%
1+e^{-4\pi Y\tanh Y}}{{\cdot} }  \nonumber
\end{equation}
Because $Y$ scales to infinity, the higher energy state completely decouples
and the total spin on the dot is quenched: $<S^{z}>\rightarrow 0$.

\section{Perturbative analysis at $T >> T_{K}$}

The Kondo Effect in metals containing magnetic impurities is responsible for
the ``anomalous'' minimum in the resistivity $\rho (T)$, as the temperature $%
T\sim T_{K}$: such a minimum in $\rho (T)$ is due to scattering of
conduction electrons off the localized magnetic impurities. Those
contributions were first worked out by Kondo \cite{kondo2}, who derived a
correction $\Delta \rho (T)\propto \ln \left( \frac{T_{K}}{T}\right) $.

As we pointed out in the introduction, a different realization of the Kondo
effect may be achieved, in a controlled way, in a quantum dot in Coulomb
Blockade (CB)-regime \cite{goldhaber}. The Kondo effect in a dot is usually
detected by measuring the linear conductance as a function of the gate
voltage by connecting two electrodes to the dot. A dot at CB exhibits
discrete energy levels. Changing the number of electrons is strongly
prevented by electrostatic charging energy. Correspondingly, the total
charge at the dot is quantized and the linear conductance is zero within
large windows of variations of the gate voltage (``CB-valleys''). By
analyzing the level structure of the dot it has been shown that, in some
special cases, the dot at CB behaves as a magnetic moment
antiferromagnetically interacting with the magnetic momenta of lead
electrons. CB-valleys at different occupation number of the dot are bounded
by resonant conduction peaks, as the chemical potential of the leads matches
the one of the dot. Those peaks usually get sharper and better defined as $T$
is lowered. However, as $T\sim T_{K}$, Kondo effect arises at the dot.
Consequently, the CB-valley between two resonant conduction peaks ``fills
in'' in a way that does not depend on the position of the Fermi level of the
leads \cite{goldhaber}. As $T\sim T_{K}$, the conductance $g(T)$ exhibits
the typical logarithmic raise \cite{silvano0}. The logarithmic raise,
however, cannot hold all the way down to $T=0$, as it must be limited by the
unitarity bound. Therefore, a different approach is required in order to
study Kondo effect in the $T=0$-unitarity limit, as we will discuss in the
next section.

In this Section we derive the Kondo Hamiltonian from the single impurity
Anderson model (eq. \ref{app1}) focusing for simplicity on the isotropic
case $J_{\perp }=J_{z}=J$, then we sketch the perturbative Renormalization
Group (RG) flow for the coupling strength of \ such a model. In general, the
lower is $T$, the more spin-flip processes become likely, which make the
running coupling constant $J$ grow. At the Kondo temperature $T_{K}$ the
perturbative analysis breaks down, that is $\nu \left( 0\right) J(T_{K})\sim
1$, so that $T_{K}$ is the characteristic scale that divides the regions of
weak and strong coupling.

\subsection{Derivation of the spin dynamics hamiltonian}

In the following we will restrict our analysis to a two-fold degenerate dot
level: in such a case the QD can be modelized as a spin-1/2 magnetic
impurity. Indeed, because the charge degree of freedom is frozen at Coulomb
Blockade we mimick the dot with its total spin $\vec{S}_{d}$ and describe
its interaction with the delocalized electrons by means of the Kondo
Hamiltonian:

\begin{equation}
H_{{\rm eff}}=H_{{\rm lead}}+H_{K}\equiv H_{{\rm lead}}-J\vec{S}_{d}{{\cdot}
}\vec{\sigma}(0),  \label{kh}
\end{equation}
where $\vec{\sigma}(0)=\sum_{kk^{^{\prime }}}\sum_{\sigma \sigma ^{^{\prime
}}}(c_{k\sigma }^{\dagger }\overrightarrow{\tau }_{\sigma \sigma ^{^{\prime
}}}c_{k^{^{\prime }}\sigma ^{^{\prime }}})$ plays the role of a spin density
of the itinerant electrons at the impurity site ($\overrightarrow{\tau }
_{\sigma \sigma ^{^{\prime }}}$ are the Pauli spin$-\met$ matrices). Its
components behave as a spin$-\met$, provided single occupancy of the site $%
x=0$ is guaranteed.

In the anisotropic Kondo model the couplings of the $x,y$ components $%
J_{\perp }$ are different from the one of the $z$ component $J_{z}$. This
effective Hamiltonian is more suitable to describe the low-$T$ physics of
the system because its dynamics shows how the system flows towards the
nonperturbative regime. In the following sections by ``low temperature''
we'll mean $T\sim T_{K}$.

As a first step we show that it is possible to derive the effective
Hamiltonian (\ref{kh}) from the one for the single impurity Anderson model
(eq. \ref{app1}), where $H_{{\rm D}}=$ $\epsilon _{d}\sum_{\sigma }n_{\sigma
}+U\sum_{\sigma \sigma ^{^{\prime }}}n_{\sigma }n_{\sigma ^{^{\prime }}}$,
by means of the Schrieffer-Wolff (SW) transformation \cite{schrieffer}. More
details can be found on the book by Hewson \cite{hewson}.

Let $\Xi $ be the space spanned by the (almost) degenerate dot's states. As
the number of electrons at the QD is fixed, the relevant degrees of freedom
can be described by an effective Hamiltonian $H_{{\rm eff}}$ acting onto $%
\Xi $ only. In order to construct $H_{{\rm eff}}$, we apply the SW
transformation to the Hamiltonian in eq. (\ref{app1}). We denote by $P$ the
projector onto $\Xi $ and by $\epsilon _{d}$ the energy of the states within
$\Xi $, so that the effective Kondo Hamiltonian is given by:

\begin{equation}
\delta H_{{\rm eff}}\approx P\;V(1-P)\frac{1}{\epsilon _{d}-H_{{\rm D}}-H_{%
{\rm lead}}}(1-P)\;V\;P.  \label{ef1}
\end{equation}

\noindent Indeed, by inserting eq. (\ref{app1}) into eq. (\ref{ef1}), we get
the result:

\begin{eqnarray}
H_{{\rm eff}} &=&H_{{\rm lead}}-\nu \left( 0\right) \sum_{\alpha \sigma }%
\frac{V_{\alpha }^{2}}{\epsilon _{d}}d_{\sigma }^{\dagger }d_{\sigma
}+\sum_{\alpha \sigma ;k,k^{^{\prime }}}V_{\alpha }^{2}\left[ \frac{1}{%
\epsilon _{d}+U}+\frac{1}{\epsilon _{d}}\right] c_{k\alpha \sigma }^{\dagger
}c_{k^{^{\prime }}\alpha \sigma }+  \nonumber \\
&&-\sum_{\alpha ,\alpha ^{^{\prime }};\sigma ,\sigma ^{^{\prime
}};k,k^{^{\prime }}}V_{\alpha }V_{\alpha ^{^{\prime }}}\left[ \frac{1}{%
\epsilon _{d}+U}-\frac{1}{\epsilon _{d}}\right] c_{k\alpha \sigma }^{\dagger
}\vec{S}_{d}{{\cdot} }\vec{\tau}_{\sigma \sigma ^{^{\prime }}}c_{k^{^{\prime
}}\alpha ^{^{\prime }}\sigma ^{^{\prime }}}  \label{ef2}
\end{eqnarray}
\noindent where $S_{d}^{z}=\sum_{\sigma }\sigma d_{\sigma }^{\dagger
}d_{\sigma }$, $S_{d}^{+}=d_{\uparrow }^{\dagger }d_{\downarrow }$ and $%
S_{d}^{-}=d_{\downarrow }^{\dagger }d_{\uparrow }$ are the impurity spin
components. Spin commutation relations are obtained if no double occupancy
is admitted.

Besides $H_{{\rm lead}}$, the first and the second term at the r.h.s. of eq.
(\ref{ef2}) represent, respectively, a renormalization of the dot's energies
and a potential scattering term; the third term is the one which induces
spin flips. The two contributions appearing in the potential scattering term
and in the spin-flip term refer to two virtual particle and hole processes,
respectively. In the first case a particle is added to the dot so that the
energy $\epsilon _{d}+U$ is involved, while in the second case a particle is
subtracted from the dot level thus paying the energy $|\epsilon _{d}|$. The
potential scattering term vanishes if the Anderson model is symmetric, as we
have assumed up to now ($\epsilon _{d}=-U/2$).

\subsection{Perturbative Renormalization Group Approach}

In this Subsection the starting point of our reasoning is, for simplicity,
the isotropic limit of Kondo hamiltonian in eq. (\ref{kh}). Scattering of
conduction electrons by the impurity produces a self energy correction, as
well as a correction to the interaction vertex. The transitions between the
states close to the Fermi level $\epsilon _{F}$ and the states within a
narrow strip of energies of width $\delta D$ near the edges of the band of
width $2D$ are associated with an high energy deficit. Such transitions are
virtual and their influence on the states near $\epsilon _{F}$ can be taken
into account perturbatively in the second order. In the {\sl poor man's
scaling } approach one includes second order corrections arising from
processes in which the electrons $k^{\prime }$ are scattered to an
intermediate state at energy $D>\epsilon _{k^{\prime \prime }}>D-\delta D$
or $-D<\epsilon _{k^{\prime \prime }}<-D+\delta D$ where $D$ is some
ultraviolet cut-off. Because they involve high intermediate energies, one
can think of including them perturbatively, so for an electron $k^{\prime }$
scattered into the final state $k$ we have:
\begin{equation}
\sum_{k^{\prime \prime }\in \gamma }<k|H_{K}|k^{\prime \prime }><k^{\prime
\prime }|\frac{1}{E-H_{{\rm lead}}}|k^{\prime \prime }><k^{\prime \prime
}|H_{K}|k^{\prime }>\sim -J^{2}\nu (0)\delta D\frac{1}{D}\vec{S}_{d}{{\cdot} }%
\sum_{kk^{^{\prime }}}\sum_{\sigma \sigma ^{^{\prime }}}c_{k\sigma
}^{\dagger }\overrightarrow{\tau }_{\sigma \sigma ^{\prime }}c_{k^{\prime
}\sigma ^{\prime }}  \nonumber
\end{equation}
where $\gamma $ is the $k^{\prime \prime }$ domain mentioned above. To
justify the second step one observes that, in the shell $\gamma $, the
operator $H_{{\rm lead}}$ can be replaced by an energy $D$ and, in
comparison to it, the eigenvalue $E$ can be put at the Fermi level $E=0$.
The result is an effective Hamiltonian acting within the band of a reduced
width $D-\delta D$, of the same form as $H_{K}$ in eq. (\ref{kh}) but with a
modified value of $J$. This gives the following correction to the previous
coupling constant in $H_{K}$: $\delta J\sim -J^{2}\nu (0)\delta D/D$ \cite
{anderson}. Successive reductions of the bandwidth by small steps $\delta D$
can be viewed as a continuous process during which the initial Hamiltonian
is trasformed to an effective low-energy Hamiltonian acting within the band
of reduced width $D-\delta D$. The evolution of the exchange amplitude $J$
during such a {\sl poor man's scaling }can be cast into the form of a flow
differential equation \cite{anderson}:
\begin{equation}
\frac{dJ}{d\ln D}=-\nu (0)J^{2}.  \label{scal0}
\end{equation}

Integration of eq. (\ref{scal0}) gives rise to the renormalization of $J$:
\begin{equation}
\frac{1}{j(D)}-\frac{1}{j(D_{0})}=\ln \frac{D}{D_{0}},  \label{scal1}
\end{equation}
where $j=J\nu (0)$, that is
\begin{equation}
j(D)=\frac{j_{0}}{1-j_{0}\ln \frac{D_{0}}{D}}.  \label{scal2}
\end{equation}
Scaling can be performed down to $D\sim k_{B}T$, also we see that, if $%
j_{0}\equiv j(D_{0})>0$ (antiferromagnetic coupling), the running coupling
constant $j$ increases. Eq. (\ref{scal1}) shows that $De^{-\frac{1}{j}}$ is
a constant of this flow, which defines an energy scale as $T_{K}$:
\begin{equation}
k_{B}T_{K}=D_{0}e^{-\frac{1}{j_{0}}}.  \label{scal3}
\end{equation}
In general, at the Kondo temperature $T_{K}$ the system has approached the
scale at which the perturbative analysis breaks down, so that for $T<T_{K}$ $%
j$ starts to diverge in the flow.

Now we discuss the conductance in such a perturbative limit. Let us take a $%
\delta (x)$ -like dot with an even barrier potential ($\Gamma _{R}=\Gamma
_{L}$). To lowest perturbative order the leading self-energy correction to $%
t^{0}(\omega +i0^{+})$ (see eqs. (\ref{gdot}), (\ref{born})) ($\omega =0$)
yields:
\begin{equation}
g=\left| \frac{\pi }{k_{o}}t^{0}\right| ^{2}=\Re e\left\{ \frac{i\pi }{k_{o}}%
t^{0}\right\} \approx \pi ^{2}(\nu (0)J)^{2},  \label{sun}
\end{equation}
where $\frac{1}{\nu (0)J}=\ln \frac{T}{T_{K}}$ is the invariant charge
defined through eq. (\ref{scal3}), so that the conductance can be written
as:
\begin{equation}
g\approx \pi ^{2}\ln ^{-2}\frac{T}{T_{K}},\text{ \ \ \ \ }T_{K}\ll T\ll D.
\label{cweak}
\end{equation}
The tail of the conductance in the perturbative limit $T\gg T_{K}$ is
logarithmic.

\subsection{One channel anisotropic Kondo model and the Toulouse limit}

In the following we rewrite the Coulomb gas approach introduced in
subsection $V.B$ focusing on the single channel anisotropic Kondo system.
Let us start from the general effective Hamiltonian:

\begin{equation}
H_{{\rm eff}}^{{\rm A}}=H_{{\rm lead}}+H_{K}^{{\rm A}}\equiv H_{{\rm lead}%
}+J_{z}S_{d}^{z}\sigma _{z}(0)+\frac{J_{\perp }}{2}\left( S_{d}^{+}\sigma
_{-}(0)+S_{d}^{-}\sigma _{+}(0)\right) .
\end{equation}

As stated in ref. \cite{yuval}, the perturbation term (the one containing $%
J_{\perp }$) has the effect of flipping the local spin at each application;
hence the problem of calculating the partition function of such a system
reduces to the evaluation of the amplitude for a succession of spin flips at
times $\tau _{1},\tau _{2},...$ and the Feynman sum over histories is the
sum over all possible numbers and positions of flips. So, we get the
expression (\ref{part1}) which can be rewritten as:

\begin{equation}
Z_{CG}=\sum_{N=0}^{\infty }\left( J_{\perp }\tau _{M}\right)
^{2N}\int_{0}^{\beta }\frac{d\tau _{2N}}{\tau _{M}}\ldots \int_{0}^{\tau
_{2}-\tau _{M}}\frac{d\tau _{1}}{\tau _{M}}[e^{+\sum_{i\neq
j=1}^{2N}(-1)^{i+j}\left( 2-\varepsilon \right) \ln \left\vert \frac{\tau
_{i}-\tau _{j}}{\tau _{M}}\right\vert }].  \label{part3}
\end{equation}
Here $\varepsilon $ is the quantity:

\begin{equation}
\varepsilon =\frac{8\delta _{AF}}{\pi }-\frac{8\delta _{AF}^{2}}{\pi ^{2}}%
\simeq 2J_{z}\tau _{M}
\end{equation}
where $\delta _{AF}$ is the scattering phase shift of antiferromagnetic sign
introduced by the $J_{z}S_{d}^{z}\sigma _{z}(0)$ term. It can be seen
clearly that the sign of $J_{z}$, or $\varepsilon $, determines whether the
coupling is ferromagnetic or antiferromagnetic, so the condition $%
\varepsilon >0$ gives the antiferromagnetic coupling.

Let us now notice that eq. (\ref{part3}) is a function of three parameters
only: $\frac{\tau }{\tau _{M}},J_{\perp }\tau _{M},\varepsilon $; we are
interested to the case $\tau \rightarrow \infty $ (low temperature). In such
a case it is well known that the ferromagnetic Kondo system has a mean spin
moment which corresponds to a long range order of the classical system: the
positive and negative charges are all bound in pairs forming dipoles all
pointing in the same direction. The phase transition occurs, at least for
small $J_{\perp }$, at the ferromagnetic-antiferromagnetic boundary point $%
\varepsilon =0$.

For such a system it is possible to derive a set of scaling equations by
renormalization of the cut-off $\tau _{M}$; these laws are exact for small $%
J_{\perp }\tau _{M}$ and $\varepsilon $. The physical picture is that of
many close pairs of flips which change slightly the mean magnetization,
located between pairs of isolated flips which are reversals of the
magnetization over a larger timescale; so, the isolated flips can be
considered as acting in a medium where the close pairs modify the mean
magnetization. This line of reasoning leads to the following scaling laws
for the \textquotedblright fugacity\textquotedblright\ $J_{\perp }\tau _{M}$
and the \textquotedblright charge\textquotedblright\ $\varepsilon $
respectively:

\begin{equation}
\frac{d\left( J_{\perp }\tau _{M}\right) }{d\ln \tau _{M}}=\frac{\varepsilon
}{2}\left( J_{\perp }\tau _{M}\right) \text{ };\text{ \ }\frac{d\varepsilon
}{d\ln \tau _{M}}=\left( 2-\varepsilon \right) \left( J_{\perp }\tau
_{M}\right) ^{2}.  \label{scaling1}
\end{equation}

Let us make a few comments on such equations. First, these are also exact
for finite $\varepsilon $, because only $J_{\perp }\tau _{M}$ need to be
small. Second, they are compatible in the isotropic case $J_{z}=J_{\perp }$,
$\varepsilon \simeq 2J_{z}\tau _{M}\simeq 2J_{\perp }\tau _{M}$ where $%
J_{\perp }\tau _{M}$ and $\varepsilon $ are small; because $J_{\perp }\tau
_{M}$ and $\varepsilon $ scale together, the isotropic Kondo system remains
isotropic at every time scale. In the anisotropic model, eqs. (\ref{scaling1}%
) become:

\begin{equation}
\frac{d\varepsilon }{d\left( J_{\perp }\tau _{M}\right) }\simeq 4\frac{%
J_{\perp }\tau _{M}}{\varepsilon }\text{ };\text{ \ }\varepsilon
^{2}-4J_{\perp }^{2}\tau _{M}^{2}=const,
\end{equation}
thus the scaling lines are a set of hyperbolas with asymptotes corresponding
to the isotropic case.


We see clearly from Fig. 3 that all ferromagnetic cases below the isotropic
one scale onto the case $J_{\perp }\tau _{M}=0$ which is an ordered one, so
the transition line coincide with the ferromagnetic case. Conversely, in the
antiferromagnetic (AF) case $J_{\perp }\tau _{M}$ and $\varepsilon $
increase starting from some values, small at will, and we always can find a
timescale for which $\varepsilon \sim 1$. Such a timescale is a crucial one
for the Kondo phenomenon, it is the Kondo temperature already introduced in
the previous sections and sets the scale at which the system behaves as it
were strongly coupled. Thus, the renormalization procedure is valid up to $%
\varepsilon \sim 1$.

Now, let us briefly focus on the so-called Toulouse limit $J_{z}=J_{\perp }$%
, $\varepsilon $ $\sim 1$; in such a case the system is equivalent to a
simple quadratic model with Hamiltonian:

\begin{equation}
H_{{\rm T}}=\sum_{k}\epsilon _{k}c_{k}^{\dagger }c_{k}+V\sum_{k}[d^{\dagger
}c_{k}+c_{k}^{\dagger }d].  \label{Toul1}
\end{equation}
Here $c_{k},c_{k}^{\dagger }$ are Fermi operators for free spinless
electrons and $d,d^{\dagger }$ are Fermi operators for a local resonant
state. The partition function corresponding to the Hamiltonian (\ref{Toul1})
is exactly equal to the one in eq. (\ref{part3}) for the case of classical
charges $\pm 1$. Such a theory corresponds to free particles and there is no
renormalization.

\section{Breakdown of perturbative approach at $T\sim T_{K}$:
strong-coupling fixed point}

In the previous Section we have derived the perturbative flow equation for
the Kondo coupling constant $J$. Integration of the renormalization group
(RG) equation shows that $J$ grows as $T$ is lowered, until the perturbative
analysis does not make any sense anymore. The question whether the RG flow
stops at some finite-$T$ scale fixed coupling or goes all the way down to $%
T=0$ has been widely debated in the literature (see \cite{hewson} for a
review on the subject). However, from numerical RG analysis and from the
exact Bethe-ansatz solution of the model \cite{tsvi} it is now clear that
the system takes no fixed points at finite $T$, but the RG flow goes all the
way down to $T=0$. So, the aim of this Section is to discuss the physics of
the Kondo system in the $T=0$-unitarity limit.

In the case of a localized spin-1/2 impurity antiferromagnetically
interacting with the spin of one type of itinerant electrons only (spin-1/2,
one channel Kondo effect), the theory of the unitary limit was first
developed by Nozi\'{e}res \cite{nozieres}. As $T$ is lowered all the way
down to 0, the flow of the coupling strength runs all the way toward an $%
\infty $-coupling fixed point. At the fixed point, the impurity spin is
fully screened and the localized magnetic moment is effectively substituted
by a spinless potential scattering center with infinite strength (which
forbids double occupancy at the impurity site). The $T\rightarrow 0$
\textquotedblleft unitarity limit\textquotedblright\ is well-described by a
Fermi liquid theory, which also allows for calculation of finite-$T$
corrections to fixed-point values of the physical quantities. Both elastic
and inelastic scattering processes provide finite-$T$ contributions to the
conductance. As we will show in the next section, the two kinds of processes
contribute at the same order, thus giving raise to corrections to the
unitary limit proportional to $\left( \frac{T}{T_{K}}\right) ^{2}$ \cite
{nozieres}.

A generalization of Kondo's original idea was done by Nozi\'{e}res and
Blandin \cite{blandin}. They showed that interaction of itinerant electrons
with magnetic impurities in metals may involve electrons with different
quantum numbers besides the spin: for instance, electrons with different
orbital angular momentum. Different orbital quantum numbers define different
\textquotedblleft channels\textquotedblright\ of interaction. Therefore,
\textquotedblleft many channel\textquotedblright\ Kondo effect may arise.
Although in the perturbative region there is no qualitative difference
between one-channel and many-channel effect, deep differences may arise in
the unitarity limit, depending on the number of channels $K$ and on the
total spin of the impurity, $S$.

In the case $2S=K$, at $T=0$ the magnetic moment at the impurity is fully
screened by electrons from the leads. The impurity forms a singlet state
with $2S$ conduction electrons and no other electrons can access the
impurity site. The system behaves exactly as if there were no impurity,
besides a boundary condition on the wavefunction of conduction electrons
which takes into account that the impurity site is ``forbidden''\ to them
\cite{nozieres}. Such a Fermi liquid state is stable against leading finite-$%
T$ corrections, as we shall see below, and corresponds to an infinite
effective coupling $\nu \left( 0\right) J$. A special case of this is the
one channel Kondo just discussed, with $S=\frac{1}{2}$ and $K=1$.

In the case $2S>K$, in the strong coupling regime a residual magnetic moment
is still present at the impurity, since there are no more conduction
electrons able to screen the localized spin. The corresponding fixed point
is again a Fermi liquid, but with a localized partially screened magnetic
moment at the impurity that is {\bf ferromagnetically} coupled to the
itinerant electrons: it again provides the stability of the local Fermi
liquid.

A very special case is the $2S<K$ one. In such a case conduction electrons
attempt to ``overscreen''\ the impurity in the strong coupling limit, that
is the resulting magnetic moment is opposite to the original one. The
coupling among the localized residual magnetic moment and the itinerant
electrons is now {\bf antiferromagnetic}. It drives the system out of the
strongly coupled regime towards a finite-coupling fixed point $J_{\ast }$
\cite{blandin}, as we clearly see in Fig. 4. At $J=J_{\ast }$ the leading
finite-$T$ interaction is given by a marginal three-particle operator which
breaks down the Fermi liquid state and generates a non-Fermi liquid behavior
in the physical quantities. We will discuss such an issue in the Section $%
VII $ of this review.


\section{The fully screened single channel Kondo case}

\subsection{Finite temperature corrections to the one-channel Kondo
conductance}

At $T=0 $ the linear response of the Kondo system to an applied voltage bias
reaches the so called unitarity limit.

The response function is the resistivity in a bulk system (magnetic
impurities in diluted alloys), while it is the conductance in a quasi
one-dimensional system as the system of our interest: a dot with applied
contacts.

The striking feature of the Kondo effect in diluted alloys is the minimum in
the resistivity at low temperature which violates the expected property: $%
d\rho /dT>0$. Indeed, well below $T_{K}$ the resistivity increases again up
to a maximum value proportional to the number of impurities $N_{i}$ per unit
volume (''unitarity limit'').

On the contrary a maximum of the conductance is expected at $T=0$ for Kondo
conductance across a quantum dot, where the unitarity limit that is reached
is $\frac{2e^{2}}{h}g_{u}$.

Being the $s-$wave scattering effectively one-dimensional like, such a
'reversed behavior' looks paradoxical. However, it is just a consequence of
the difference between the 3-d and 1-d impurity scattering, as we explain
here. In order to illustrate the difference, it is enough to note that both
facts stem from the main feature of Kondo impurity scattering: the formation
of a resonance at the Fermi level due to many-body effects, which implies
that the phase shift reaches the value $\delta =\pi /2$ and that scattering
is resonant at the impurity. This produces different results in the two
cases:

\begin{itemize}
\item[ 3-d:]  a spherical $s-$ wave is diffracted from the impurity with
maximum amplitude at the resonance, what increases the flux propagating
backward and enhances resistivity.

\item[ 1-d :]  resonant scattering coincides with resonant trasmission in
this case, what implies the vanishing of backward reflection and enhances
the conductance.
\end{itemize}

Temperature corrections are twofold. One is due to the energy dependence of
the phase shift close to the resonance $\delta ^{0}(\epsilon )=\pi
/2-a\epsilon ^{2}$ and to the fact that energies close to the Fermi surface
are sampled, because the Fermi functions are not step-like at finite $T$.
The second one is due to inelastic processes which produce transitions from
the singlet ground state to the excited states. The latter can be accounted
for with an expansion in inverse powers of the singlet binding energy \cite
{nozieres}. We include a simplified approach to the problem which rests on
the Fermi liquid nature of the excitation spectrum in Appendix B \cite
{nozieres} \cite{ludaff}\cite{costi}. It is found that corrections to the
conductance are quadratic in temperature, as it is usual in the Fermi liquid
theory:
\begin{equation}
g=g_{u}\left[ 1-\left( \frac{\pi T}{T_{K}}\right) ^{2}\right] ,\text{ \ \ \
\ }T\ll T_{K}.  \label{cstrong}
\end{equation}

The weak-coupling ($T\gg T_{K}$) and the strong-coupling ($T\ll T_{K}$)
asymptotes of the conductance, eqs. (\ref{cweak}) and (\ref{gdot}), have a
very different structure but, since the Kondo effect is a crossover
phenomenon rather than a phase transition \cite{nrg}\cite{tsvi}, the
dependence $g\left( T\right) $ is a smooth function \cite{costi} throughout
the crossover region $T\sim T_{K}$:
\begin{equation}
g=g_{u}f\left( \frac{T}{T_{K}}\right) .
\end{equation}
The universal function $f\left( x\right) $, as found by resorting to
numerical renormalization group (NRG) in refs. \cite{costi}\cite{costi1}, is
plotted in Fig. 5. It interpolates between $f\left( x\gg 1\right) =\frac{%
3\pi ^{2}}{16}\left( \ln x\right) ^{-2}$ and $f\left( 0\right) =1$.

\subsection{ The Kondo resonance in a magnetic field}

A small magnetic field $B$ lifts the degeneracy of the spin states at the
impurity.This produces a splitting of the resonance and a change of its
shape which has been numerically studied mostly via NRG \cite{costi1}. The
splitting of the peak increases linearly with $B$ and is $\Delta =2\mu _{B}B$%
, {\sl i.e.} twice the Zeeman spin splitting.

This can be understood easily if one considers the particle and hole virtual
occupations which mediate the Kondo interaction.

Let us consider the symmetric case $\epsilon _{d}=-U/2$. The dot states with
$N$ electrons and one unpaired spin are $N\uparrow $ and $N\downarrow $
corresponding to $\epsilon _{d}+B/2$ ($\mu _{B}=1$) and $\epsilon _{d}-B/2$,
respectively. In presence of spin splitting two particle processes are
allowed: $p_{\sigma }$ ($p_{-\sigma }$) in which a spin $-\sigma $ is added
to the state $N\sigma $ and subsequently a spin $-\sigma $ is removed with
the energy balance:

$p_\downarrow \to \delta E (N\uparrow, +\downarrow ) + \delta E (N+1 ,-
\uparrow ) = U + (-U-B ) = -B $

$p_\uparrow \to \delta E (N\downarrow, +\uparrow ) + \delta E (N+1 ,-
\downarrow ) = (U + B) + (-U ) = B $.

Similarly the hole processes have an energy balance:

$h_\uparrow \to \delta E (N\uparrow, -\uparrow ) + \delta E (N-1, +
\downarrow ) = (-\epsilon _d - B/2) + (\epsilon _d -B/2 ) =- B $

$h_{\downarrow }\rightarrow \delta E(N\downarrow ,-\downarrow )+\delta
E(N-1,+\uparrow )=(-\epsilon _{d}+B/2)+(\epsilon _{d}+B/2)=B$.

This implies that there are two peaks in the spectral density at $\omega =
\pm B $ corresponding to the $p_\downarrow ,h_\uparrow $ and $p_\uparrow
,h_\downarrow $ spin flip processes, respectively.

The value at the Fermi energy of the spectral function is related to the $B$%
-dependent phase shift in the Fermi liquid picture by the Friedel sum rule:
\begin{equation}
-\pi \nu \left( 0\right) \Im m\left( t_{\sigma }\left( \omega
=0,T=0,B\right) \right) =sin^{2}\delta _{\sigma }(B).
\end{equation}
The Bethe Ansatz solution of the problem relates the phase shift at the
Fermi level to the magnetization of the impurity: $\delta _{\sigma }(B)=%
\frac{\pi }{2}[1-2M_{d}(B)]$ \cite{andrei}. Hence a reduction of the peak
height with increasing of the magnetic field follows.

It has also been argued that the renormalizability of the Kondo problem
could break down when $B$ exceeds some critical value related to the Kondo
temperature, because of the back action of the induced conduction electron
polarization cloud on the impurity leading to a broken symmetry state with $%
<S_{z}>\neq 0$ \cite{ovchinnikov}.

\subsection{Crossing of the dot levels in a magnetic field and enhancement
of the Kondo conductance}

Conventional Kondo resonant transmission requires a magnetic moment to be
present on the dot. Usually dots with an even number of electrons are in a
singlet state, while dots with an odd number of electrons have an unpaired
spin and have a doublet GS. Hence, there is a parity effect: CB conduction
valleys with $N=even$ do not display the Kondo conductance while those with $%
N=odd$ do. An exception to this rule occurs at zero magnetic field when
Hund's rule applies. This was found experimentally in vertical QD e.g. at $%
N=6$. The GS of the isolated dot has $S=1$ (triplet) \cite{tarucha}. An
underscreened Kondo effect is expected at this point and the GS of the
interacting system becomes a doublet. By applying a weak magnetic field
orthogonal to the dot ($B=0.22Tesla$), a transition of the GS from triplet
to singlet (T- S) has been found. The single particle energy levels become
angular momentum dependent and Hund's rule no longer applies. Close to the
crossing a remarkable enhancement of Kondo coupling with increase of the
Kondo temperature was experimentally found. Indeed, scaling shows a non
universal critical temperature when the interplay of the fourfold degenerate
states ($|SS_{z}>$ with $S=0,1$) is included \cite{eto}.

An extended and unified approach to the problem can be found in \cite
{pustil1}. A minor difference is the fact that they consider an in-plane
magnetic field as the source of the crossing which could produce Zeeman spin
splitting (ZSS) of the single particle states. The four dot states are
mapped onto a two impurities Kondo model (2IKM) \cite{twoimp} with spin $%
\overrightarrow{S}_{1},\overrightarrow{S}_{2}$. However they are coupled by
a potential term $Vn\rho _{nn}(0)\overrightarrow{S}_{1}{\cdot} \overrightarrow{%
S}_{2}$ and an exchange term $iIns_{\overline{n}n}(0){\cdot} \overrightarrow{S}%
_{1}{\times} \overrightarrow{S}_{2}$. Here $\rho ,\sigma $ are the charge and
spin density of the conduction electrons with $n$ and $\overline{n}\neq n$
labeling two different orbital states. Because these terms violate the
invariance under particle-hole transformation, the 2IKM cannot flow by
scaling to the non Fermi-liquid fixed point, which is known to be a
remarkable feature of the model. In the case of large ZSS $\Delta $ the RG
flow terminates at $D\approx \Delta $. Two of the four states are ruled out
in the flow and conduction electrons couple to one single effective spin $%
1/2 $ with one extra unusual term in the effective Hamiltonian which is a
Zeeman term for the conduction electrons.

Kondo conductance can take place in a dot with an even number of electrons
also in a strong vertical magnetic field \cite{arturo2}. Orbital effects
induced by B can produce the reversed transition from the singlet to the
triplet state (S-T). Indeed, a vertical magnetic field on an isolated dot
favors transitions to higher spin states \cite{wagner}. In this case the ZSS
is anyhow sizeable and the crossing involves the singlet state and the
component of the triplet state lowest lying in energy. In a vertical
geometry with cylindrical symmetry orbital angular momentum $m$ and $z-$
spin component $\sigma $ are good quantum numbers. In particular because the
singlet state has total angular momentum $M=0$ and the triplet state
involved in the crossing has $M=1$, only a $(m=0\downarrow )$ electron can
enter the dot when it is in the triplet state. On the contrary, only an $%
(m=1\uparrow )$ electron can enter the dot, if it is in the singlet state.
This implies that there is one single channel of conduction electrons
involved in cotunneling processes, with orbital and spin degrees of freedom
locked together. Again, a residual effective spin $1/2$ survives at the dot,
while $N$ is even \cite{noi}. This is another striking manifestation of the
spin-charge separation that occurs at the Kondo fixed point. Usual Kondo
coupling leads to $N=odd$ together with $S=0$. In the complementary
situation here described, it is $N=even$ and $S=1/2$.

\section{The overscreened two channel Kondo case}

A two channel Kondo behavior has been invoked in an experiment by Ralph and
Buhrman \cite{ralph} on clean Cu point contacts, defects in the metal that
can be described by two level systems (TLS). The TLS could tunnel between
the even and odd state with the assistance of the conduction electrons.
Their physical spin is not involved in the scattering, so that two channels
are available \cite{zawa}. It is still unclear whether the Kondo temperature
can be large enough so that any effect of the Kondo correlation can be
measured \cite{boris}. However, these experiments have triggered renewed
interest in two channel Kondo conductance. The temperature and voltage
dependence of the conductivity have been numerically calculated within the
``Non Crossing Approximation`` (NCA) \cite{hettler} and found to be
consistent with a scaling Ansatz motivated by the Conformal Field Theory
(CFT) solution of the problem \cite{ludaff}. Since then, no other
experimental proof of two channel Kondo effect in impurities has been
produced. The result for the imaginary part of the transmission $t$ is:
\begin{equation}
{\Im }mt(\omega )\sim \sqrt{\frac{\omega }{T_{K}}}.  \label{nflres}
\end{equation}
\noindent From this result, we get a $\sqrt{T/T_{K}}$ temperature dependence
of the conductance as $T\rightarrow 0$, which is a clear signature of the
breakdown of the Fermi liquid \cite{ludaff} (see Fig. 6).

In order to emphasize the deep difference between single-channel and
many-channel Kondo effect in the $T=0$ limit we just mention here that the
imaginary part of the proper self-energy, close to the Fermi surface,
behaves as $\Im m\Sigma (k,\omega )\propto C_{k}\omega ^{2}$ (where the
chemical potential $\mu =0$ is taken as reference energy) in the Fermi
liquid case and in the single channel Kondo problem. On the contrary, it
behaves as $\Im m\Sigma (k,\omega )\propto C_{k}^{\prime }\omega ^{\met}$ in
the two channel ''overscreened''\ Kondo problem. In the following we refer,
for simplicity, to the two channel ''overscreened''\ case. Through the
Kramers-Kronig relation
\begin{equation}
\frac{\partial }{\partial \omega }\Re e\Sigma (k,\omega )=-\frac{1}{\pi }%
P\int_{-\infty }^{\infty }d\omega ^{\prime }\frac{\partial _{\omega }\Im
m\Sigma (k,\omega ^{\prime })}{\omega ^{\prime }-\omega }
\end{equation}
we see that $\frac{\partial }{\partial \omega }\Re e\Sigma (k,\omega
)|_{\omega \rightarrow 0}$ is finite in the first case, at the Fermi
surface, on the contrary it has a power law divergence in the second case.
It follows that the quasiparticle pole residue $z_{k}=[1-\partial _{\omega
}\Sigma (k,\omega )]^{-1}|_{\omega \rightarrow 0}$ vanishes as a power law
in the second case at the Fermi surface. Luttinger \cite{luttinger} showed
that, to all orders of perturbation theory in the interaction, the imaginary
part of the proper self-energy behaves as $\Im m\Sigma (k,\omega )\propto
C_{k}\omega ^{2}$ ($C_{k}>0$) close to the Fermi surface, which implies an
infinite lifetime for the quasiparticles at the Fermi surface. The Fermi
surface is sharp and well defined. These are the foundation stones of the
normal Fermi liquid Theory and they are invalidated in the spin-$\met$ two
channel Kondo case.

In this Section we focus, in particular, on two channel spin-1/2 Kondo
effect in both the perturbative region and the unitarity limit. We describe
the scaling perturbative approach and the bosonization ($T\sim 0$)
technique, respectively. An attempt to extend the Anderson-Yuval approach of
Section $IV.C$ to the two channel Kondo case can be found in \cite{fabrizio}.


\subsection{\protect\bigskip Perturbative analysis at $T>> T_{K}$}

The starting point of our reasoning is the effective Hamiltonian given in
eq. (\ref{ef2}), which we will take in the isotropic limit. In the following
we will restrict our analysis to a two-fold degenerate dot level. In this
case, as we pointed out, the QD can be modelized as a spin-1/2 magnetic
impurity, whose spin is given by:

\[
S_{d}^{a}=\frac{1}{2}d_{\gamma }^{\dagger }\tau _{\gamma \gamma ^{^{\prime
}}}^{a}d_{\gamma ^{^{\prime }}}
\]
\noindent (that $\vec{S}_{d}$ is a spin-1/2 comes out from the identity $%
\vec{S}_{d}^{2}=3/4$, valid in the case of single-occupancy for the dot's
level).

In this case and for a generic number of channels for the itinerant
electrons, eq. (\ref{ef2}) takes the form:

\begin{equation}
H_{K}=J\sum_{a}\left( \sum_{\gamma \gamma ^{^{\prime }}}(d_{\gamma
}^{\dagger }\frac{\tau _{\gamma \gamma ^{^{\prime }}}^{a}}{2}d_{\gamma
^{^{\prime }}})\sum_{kk^{^{\prime }}}\sum_{\sigma \sigma ^{^{\prime
}};\alpha }(c_{k\sigma \alpha }^{\dagger }\frac{\tau _{\sigma \sigma
^{^{\prime }}}^{a}}{2}c_{k^{^{\prime }}\sigma ^{^{\prime }}\alpha })\right)
\label{heftru}
\end{equation}
\noindent where $\alpha \in (1,..,K)$ is the channel index and the constant $%
J$ is taken as a perturbative parameter ($>0$). Infrared divergent diagrams
provide a flow of $J$ as a function of $T$. We are now going to derive the
perturbative $\beta $-function at third order in $J$. At finite $T$ the
Green function in Fourier space will depend on the momentum of the particles
and on the Matsubara frequencies $\omega _{m}=\frac{2\pi }{\beta }(m+\frac{1%
}{2})$ (for fermions). In our case, the Green function for the lead
electrons is given by:

\begin{equation}
G_{\sigma \sigma ^{^{\prime }};\alpha \alpha ^{^{\prime }}}(i\omega _{m},k)=%
{\bf FT}\left\{ \langle \hat{T}[c_{\sigma \alpha }(\tau ,k)c_{\sigma
^{^{\prime }}\alpha ^{^{\prime }}}^{\dagger }(0,k)]\rangle \right\} =\frac{%
\delta _{\sigma \sigma ^{^{\prime }}}\delta _{\alpha \alpha ^{^{\prime }}}}{%
i\omega _{m}-v_{F}k}
\end{equation}
where ${\bf FT}$ stands for 'Fourier transform' and $\hat{T}$ is the time
ordering operator, while the Green function for the $d$-fermion is:
\begin{equation}
{\cal G}_{\gamma \gamma ^{^{\prime }}}(i\omega _{m})={\bf FT}\left\{ \langle
\hat{T}[d_{\gamma }(\tau )d_{\gamma ^{^{\prime }}}^{\dagger }(0)]\rangle
\right\} =\frac{\delta _{\gamma \gamma ^{^{\prime }}}}{i\omega _{m}}.
\end{equation}
\noindent The interaction vertex determined by $H_{{\rm eff}}$ is:

\begin{equation}
V_{\sigma \sigma ^{^{\prime }};\gamma \gamma ^{^{\prime }}}^{\alpha \alpha
^{^{\prime }}}(\{i\omega _{m}^{(j)}\})=\delta _{\alpha \alpha ^{^{\prime }}}%
\frac{J}{4}\tau _{\gamma \gamma ^{^{\prime }}}^{a}\tau _{\sigma \sigma
^{^{\prime }}}^{a}\delta (\omega _{m}^{(1)}+\omega _{m}^{(2)}-\omega
_{m}^{(3)}-\omega _{m}^{(4)}).
\end{equation}
\noindent The one-loop structure of the theory provides a renormalization to
the interaction vertex, that is, to the coupling constant, by means of the
two diagrams drawn in Fig. 7. The sums over Matsubara frequencies can be
calculated by using the standard techniques described, for example, in \cite
{fetter}. In the low-energy limit for the electrons from the leads (that is,
if only excitations about the Fermi level are taken into account), the sum
of the two diagrams is given by:

\begin{equation}
{\cal D}_{1}+{\cal D}_{2}\approx -i\frac{J^{2}}{8}\delta _{\gamma \gamma
^{^{\prime }}}\{[\delta _{\sigma \sigma ^{^{\prime }}}\delta _{\alpha \alpha
^{^{\prime }}}+\tau _{\sigma \sigma ^{^{\prime }}}^{a}\tau _{\alpha \alpha
^{^{\prime }}}^{a}]+[-\delta _{\sigma \sigma ^{^{\prime }}}\delta _{\alpha
\alpha ^{^{\prime }}}+\tau _{\sigma \sigma ^{^{\prime }}}^{a}\tau _{\alpha
\alpha ^{^{\prime }}}^{a}]\}2\nu \left( 0\right) \ln (\frac{D}{k_{B}T})
\end{equation}
\noindent ($D$ is an ultraviolet cutoff, identified with the width of the
conduction band). The corresponding renormalization to the coupling constant
$J$ is easily worked out and is given by:

\begin{equation}
\Delta ^{(2)}J(T,D)=2\nu \left( 0\right) J^{2}\ln (\frac{D}{k_{B}T}).
\label{ren1}
\end{equation}
\noindent

A careful analysis of the vertex renormalization at third order in $J$
reveals that several third-order diagrams are already taken into account by
the scaling equation generated by the second-order vertex correction, as
discussed in \cite{zawa2}. The only ``new''\ contribution comes from the
``non-parquet''\ diagram shown in Fig. 8. Because of the loop over the
fermions from the contacts, such a contribution carries an overall factor of
$K$, that is, the diagram is proportional to the number of channels and the
correction to the coupling constant up to third-order in $J$ is:

\begin{equation}
\Delta ^{(3)}J=(2\nu \left( 0\right) J^{2}-2K\left( \nu \left( 0\right)
\right) ^{2}J^{3})\ln (\frac{D}{k_{B}T}).  \label{ren2}
\end{equation}
\noindent Integration of the renormalization group equation for $J$ provides:

\begin{equation}
-\frac{1}{2j}+\frac{1}{2j_{0}}+\frac{K}{2}\ln \left[ \frac{j}{j_{0}}\left(
\frac{1-Kj_{0}}{1-Kj}\right) \right] =x-x_{0}  \label{flow1}
\end{equation}
\noindent where $j=\nu \left( 0\right) J$ and $x=\ln (\frac{D}{k_{B}T})$.

It is seen from eq. (\ref{flow1}) that the fixed point is reached at an
intermediate coupling $j^*$. However, the perturbative RG analysis is not
conclusive here, because even in the ordinary, single channel, Kondo model
an artificial intermediate coupling fixed point is produced when the
perturbative RG equations are expanded to 3rd order \cite{zawa2}.

It is straightforward to infer the Kondo temperature $T_{K}$ from eq. (\ref
{flow1}). Usually $T_{K}$ is defined as the temperature scale at which $j$
becomes ${\cal O}(1)$. Following such a criterion, at $T=T_{K}$ we can
neglect $1/j$ compared to $1/j_{0}$ and obtain the following approximate
formula for $T_{K} $:

\begin{equation}
k_{B}T_{K}\approx D(j_{0})^{\frac{K}{2}}e^{-\frac{1}{2j_{0}}}.
\label{tkondo}
\end{equation}
\noindent Eq. (\ref{tkondo}) is quite general, in that it provides the value
for the crossover temperature for any number of channels $K$. This proves
that the way the system approaches the scale at which the perturbative
analysis breaks down does not depend on the number of channels, except for a
redefinition of the Kondo temperature. Hence, logarithmic divergencies are
expected when approaching to the $T_{K}$, no matter on what the number of
channels is. In the next subsection we will analyze the $T=0$ behavior of \
such a system, and will see that it is dramatically different from the one
channel case for what concerns the fixed-point properties.

\subsection{Analysis at $T\sim 0$}

Several techniques have been applied in order to get informations about
physical quantities around the fixed point in the Kondo regime in such a
limit. Finite-$T$ corrections have been derived by means of bosonization
techniques \cite{guinea}\cite{emery}, of Bethe-ansatz like exact solutions
\cite{tsvi} and of Conformal Field Theory \ (CFT) techniques \cite{ludaff}.
Now, CFT approach is extremely effective in calculating finite-$T$
corrections, Wilson ratios and other exact results concerning Green
functions \cite{ludaff}, but its starting point that charge, spin and
\textquotedblright flavor\textquotedblright\ quantum numbers are enough to
identify a primary fermionic field in the theory leads to an inconsistency:
the corresponding on-shell $S$ matrix comes out to be $0$. The solution to
such an \textquotedblright unitarity paradox\textquotedblright\ has been
suggested by Ludwig and Maldacena \cite{maldacena}, who introduced a fourth
\textquotedblright spin-flavor\textquotedblright\ quantum number. So, while
at the unitarity limit charge, spin and flavor do not change upon scattering
off the impurity, the spin-flavor changes, giving rise to one more
scattering channel. Such an extra quantum number allows for an off-diagonal
on-shell $S$ matrix, therefore the unitarity paradox simply means that the
diagonal elements of the $S$ matrix with respect to the spin-flavor number
are $0$. However, a theory of the unitarity limit, needed in order to
compute, for instance, within an unified framework transport properties with
finite-$T$ corrections, is not yet fully developed. In this subsection we
briefly sketch the first steps in order to derive an appropriate scattering
potential a' la Nozi\'{e}res, in the case of two channel spin-1/2
overscreened Kondo effect. We follow an approach equivalent to the one by
Tsvelick \cite{tsvelick}, that is the introduction of a regularization
procedure able to move the fixed point toward infinite-coupling. Then, we go
through a bosonization - refermionization in order to account for the
scattering processes with changing of the spin-flavor. The bosonization
procedure allows us to split the degrees of freedom involved in our problem:
in this way we will show that the Kondo interaction involves only the spin
and spin-flavor degrees of freedom while the charge and flavor ones are
fully decoupled. This remark is a crucial one because it makes possible to
derive a $S$ matrix in the unitarity limit which comes out to be diagonal in
any quantum number, but the spin-flavor; such a $S$ matrix describes the
whole system of dot and contacts.

In the following we discuss the case $K=2,S=1/2$, so we have two flavors of
conduction electrons from an effectively one-dimensional conductor which
interact with a localized spin$-1/2$ magnetic impurity.

Let $c_{\alpha \sigma }(x)$ be the lead electron operators ($\sigma
=\uparrow ,\downarrow $ is the spin index, $\alpha =1,2$ is the flavor
index). A lattice-model complete Hamiltonian is written as:

\begin{eqnarray}
H^{2Ch} &=&-t\sum_{x}[c_{\alpha \sigma }^{\dagger }(x)c_{\alpha \sigma
}(x+a)+c_{\alpha \sigma }^{\dagger }(x+a)c_{\alpha \sigma }(x)]  \nonumber \\
&&-\mu \sum_{x}c_{\alpha \sigma }^{\dagger }(x)c_{\alpha \sigma }(x)+J%
\overrightarrow{S}_{d}{\cdot} \lbrack \vec{\sigma}_{1}(0)+\vec{\sigma}%
_{2}(0)]
\end{eqnarray}
\noindent where $\vec{\sigma}_{\alpha }(x)=\frac{1}{2}c_{\alpha \sigma
}^{\dagger }(x)\vec{\tau}_{\sigma \sigma ^{^{\prime }}}c_{\alpha \sigma
^{^{\prime }}}(x)$ and $a$ is the lattice spacing.

Now, we can linearize the dispersion relation around the Fermi surface and
introduce two chiral fields $c_{\pm ,\alpha \sigma }(x)$. Even and odd
parities can be introduced to obtain fields with the same chirality, $\phi
_{\alpha \sigma }^{e}$ and $\phi _{\alpha \sigma }^{o}$, so odd parity fully
decouples from the interaction Hamiltonian. In order to properly deal with
the interacting fields, we will bosonize $\phi _{\alpha \sigma }^{e}$; in
particular, we define four bosonic fields $\Psi _{\alpha \sigma }$ in terms
of which it is possible to express densities of the remarkable physical
quantities. To this end we are led to construct the four bosonic fields $%
\Psi _{X}$ ($X=$ch,sp,fl,sf) for the charge, spin, flavor and spin-flavor
degrees of freedom \cite{emery}\cite{maldacena}\cite{mfab} as linear
combinations of the previous ones.

In this way it is possible to realize two ``inequivalent'' representations
of the fields $\phi _{\alpha \sigma }^{e}$ in terms of the fields $\Psi _{X}$
($X=$ch,sp,fl,sf), $\phi _{\alpha \sigma ;I}^{e}$ and $\phi _{\alpha \sigma
;II}^{e}$, given by:

\begin{eqnarray}
\phi _{\alpha \sigma ;I}^{e}(x) &=&\eta _{\alpha \sigma }:e^{-\frac{i}{2}%
[\Psi _{{\rm ch}}+\sigma \Psi _{{\rm sp}}+\alpha \Psi _{{\rm fl}}+\alpha
\sigma \Psi _{{\rm sf}}](x)}:  \nonumber \\
\phi _{\alpha \sigma ;II}^{e}(x) &=&\xi _{\alpha \sigma }:e^{-\frac{i}{2}%
[\Psi _{{\rm ch}}+\sigma \Psi _{{\rm sp}}+\alpha \Psi _{{\rm fl}}-\alpha
\sigma \Psi _{{\rm sf}}](x)}:  \label{39}
\end{eqnarray}
\noindent where $\eta $ and $\xi $ are suitable Klein factors. Notice that
the two fields differ only in the spin-flavor quantum number but such a
difference is a crucial one.

Now we are ready to rewrite the two channel Kondo interaction Hamiltonian in
bosonic coordinates $\Psi _{X}$ as:

\begin{eqnarray}
H_{K}^{2Ch} &=&J\left\{ S_{d}^{+}:e^{-i\Psi _{{\rm sp}}(0)}::\cos (\Psi _{%
{\rm sf}}(0)):+S_{d}^{-}:e^{i\Psi _{{\rm sp}}(0)}::\cos (\Psi _{{\rm sf}%
}(0)):+S_{d}^{z}\frac{L}{2\pi }\frac{d\Psi _{{\rm sp}}(0)}{dx}\right\}
\nonumber \\
&=&J\overrightarrow{S}_{d}{{\cdot} }[\vec{\Sigma}_{g}(0)+\vec{\Sigma}_{u}(0)]
\label{bosi}
\end{eqnarray}
where the spin densities $\vec{\Sigma}_{A,B}(x)$ are given by:

\begin{eqnarray}
\Sigma _{g}^{\pm }(x) &=&:e^{\pm i[\Psi _{{\rm sp}}+\Psi _{{\rm sf}}](x)}:%
\hspace*{1cm}\Sigma _{g}^{z}(x)=\frac{L}{4\pi }\frac{d}{dx}[\Psi _{{\rm sp}%
}+\Psi _{{\rm sf}}](x),  \nonumber \\
\Sigma _{u}^{\pm }(x) &=&:e^{\pm i[\Psi _{{\rm sp}}-\Psi _{{\rm sf}}](x)}:%
\hspace*{1cm}\Sigma _{u}^{z}(x)=\frac{L}{4\pi }\frac{d}{dx}[\Psi _{{\rm sp}%
}-\Psi _{{\rm sf}}](x).  \label{eq43}
\end{eqnarray}

Both $\vec{\Sigma}_{g}(x)$ and $\vec{\Sigma}_{u}(x)$ are $SU(2)$ spin
current operators, so the vector space is made of \ the bosonic vacuum $%
|bvac\rangle $ and the bosonic spin$-\frac{1}{2}$ spinors at a point $x$:

\begin{eqnarray}
|\sigma _{g}\rangle &\equiv &:e^{\sigma \frac{i}{2}[\Psi _{{\rm sp}}+\Psi _{%
{\rm sf}}](x)}:|bvac\rangle  \nonumber \\
|\sigma _{u}\rangle &\equiv &:e^{\sigma \frac{i}{2}[\Psi _{{\rm sp}}-\Psi _{%
{\rm sf}}](x)}:|bvac\rangle .  \label{eq44}
\end{eqnarray}
\noindent No triplet adding up $g$ and $u$ spin species together can occur
because, given $\vec{\Sigma} _{g,u}=\int_{-L/2}^{L/2}dx:\vec{\Sigma}%
_{g,u}(x) $, we have:
\begin{equation}
\vec{\Sigma}_{g}|\sigma _{u}\rangle =0\hspace*{1cm}\vec{\Sigma}_{u}|\sigma
_{g}\rangle =0,  \label{eq49}
\end{equation}
\noindent that is, if at a point $x$ the spin density associated to $\vec{%
\Sigma}_{g}$ is $\neq 0$, then, at the same point, the spin density
associated to $\vec{\Sigma}_{u}$ is $=0$, and vice versa; this does not
allow for overscreening. Such a statement is a crucial one, indeed it
corresponds to a particular regularization scheme \cite{tsvelick} able to
move the finite-coupling fixed point corresponding to the unitarity limit
down to $J =+\infty $. At this infinitely-strongly coupled fixed point the
impurity spin will be fully screened in a localized spin singlet. In
principle, the system might lay within any linear combination mixing the two
representations $g$, $u$, of the form:

\begin{equation}
|GS\rangle _{\mu } \left | _{x=0} \right . =|\Uparrow \rangle \otimes \frac{1%
}{2}\biggl (|\downarrow _{g}\rangle +\mu |\downarrow _{u}\rangle \biggr )%
\biggr |_{x=0}-|\Downarrow \rangle \otimes \frac{1}{2}\biggl (|\uparrow
_{g}\rangle +\mu |\uparrow _{u}\rangle \biggr ) \biggr |_{x=0}  \label{symm1}
\end{equation}
where $|\Uparrow \rangle $, $|\Downarrow \rangle $ are the two impurity
states with opposite spin polarizations. Then we search for the two
independent linear combinations that do not change upon scattering of lead
electrons; it can be shown that such combinations correspond to the values $%
\mu =\pm i$.

Now, in the fixed point limit, we ''refermionize''. Physical states mix both
representations, $I$ and $II$. Scattering by the impurity states should
conserve all physical quantum numbers. It can be shown that the elastic
scattering in the unitarity limit swaps the two inequivalent representations
$I-II$ of the lead electrons. Correspondingly, the impurity absorbs/emits
one spin-flavor quantum. The even parity $S$ matrix has the following
representation in the $I-II$ space:
\begin{equation}
{\bf S}^{0}{\bf (}\omega =0)=\left(
\begin{array}{cc}
0 & -i \\
i & 0
\end{array}
\right) \equiv \left(
\begin{array}{cc}
0 & e^{-i\frac{\pi }{2}} \\
e^{i\frac{\pi }{2}} & 0
\end{array}
\right) .  \label{matS}
\end{equation}
According to eq. (\ref{matS}) the phase shifts induced by the scattering are
$\delta ^{I\text{ }II}{\bf (}\omega =0)=-\frac{\pi }{4}$ , $\delta ^{II\text{
}I}{\bf (}\omega =0)=\frac{\pi }{4}$, while in the case of one channel spin$%
-1/2$ Kondo effect the phase shift was $\delta ^{0}=\frac{\pi }{2}$.
Finally, the conductance can be easily obtained by the Landauer formula
using eq. (\ref{matS}) and ${\bf S}^{1}=-{\bf 1}$:
\begin{equation}
g=Tr\left\{ {\bf T}\right\} =Tr\left\{ \left| \frac{1}{2}\sum_{l}S^{l}%
\right| ^{2}\right\} =\frac{1}{2}Tr\left(
\begin{array}{cc}
1 & i \\
-i & 1
\end{array}
\right) =1.
\end{equation}

\noindent Incidentally, we point out that the groundstate degeneracy always
decreases under renormalization, which is the content of ''${\rm g}$%
-theorem'' \cite{ludaff}. This leads to a zero temperature entropy for the
impurity given by ${\rm S}_{imp}\left( 0\right) =\frac{1}{2}\ln 2$.

Temperature corrections to this result have also been calculated \cite
{tsvelick} by using real fermion coordinates (Majorana fermions) $\psi
^{a}(x)(a=1,2,3)$, which obey the anticommutation relations $\{\psi
^{a}(x),\psi ^{b}(y)\}=\delta ^{ab}\delta (x-y)$, describing the relevant
coordinates only. This corresponds as well to a regularization scheme where
the fixed point has been moved to $J=\infty $. The $\sqrt{T}$ behavior is
recovered, as metioned in the introduction to this Section (see eq.(\ref
{nflres})).

\subsection{Can we reach the two channel Kondo fixed point in a quantum dot ?%
}

A possible experimental realization of the two-channel Kondo fixed point in
a quantum dot has been recently proposed \cite{arturo3}: exact
diagonalization results of a vertical quantum dot with five electrons show
that it can be tuned, by means of a strong external magnetic field, at the
degeneracy point between energy levels with $S=1/2$, but with different
orbital angular momentum. Vertical cylindrical contacts provide single
particle energy subbands labeled by the cross-sectional angular momentum and
the $k$ vector of the incoming/outgoing electron. Appropriate tuning of the
electron density in the contacts offers the chance of including two electron
channels only. The advantage of this setting is that no exchange coupling
can take place between the channels due to symmetry, so that they act as
totally independent. Selection rules due to the cylindrical symmetry enforce
angular momentum $m$ conservation and spin component $\sigma $ conservation
in the cotunneling processes. For the special setting of the proposed device
only a hole $(h)$ process for $\downarrow -$spin and a particle process for $%
\uparrow -$spin are allowed. They differ in the sign of the potential
scattering but this difference is inessential. In fact, a particle-hole
transformation on the fermion fields of the $\uparrow $ channel only
reverses the unwanted sign without affecting the exchange coupling. This
puts both channels on an equal footing and points to an ``orbital'' Kondo
coupling where the spin acts as the label for the channel. At this stage the
dot plays the role of an effective spin $3/2$ impurity, interacting with two
channels. So, as it stands the system would flow to an underscreend
situation in lowering the temperature. However, provided that Zeeman spin
splitting $\Delta $ is larger than the Kondo temperature for the
underscreened fixed point $T_{K}^{u}$, there is a crossover temperature $%
T^{c}$ at which two of the four $S=1/2$ states are ruled out of the
cotunneling processes, thus allowing for an effective spin $1/2$ two channel
Kondo flow.

Detecting the effect requires an appropriate control of the hybridization of
the dot with the contacts (so that $\Delta >>T_{K}^{u}$) and a proper tuning
of the gate voltage $V_{g}$. This allows for fine tuning of the exchange
couplings between the dot and the two channels. It is well known that the
hardest condition to be realized is total equivalence of the channels in the
coupling. Were this not the case, the system would prefer one of the two
channels as the dominant one and flow to a more conventional one-channel
Kondo state. By tuning $V_{g}$ one can cross the point where the two
channels are equally coupled which allows for scaling towards the
two-channel fixed point. The measured conductance $g\left( T\right) $ {\sl %
vs }$T$ will exhibit quadratic behavior at low temperature except for a
crossover to a square root behavior, at the appropriate $V_{g}$ value which
makes the two channels perfectly equivalent (see Fig. 6). The delicate point
in this proposal is that the requirement of full cylindrical symmetry of the
device is essential.

Another setup for measuring the two channel Kondo conductance has been
proposed recently \cite{oreg}. This proposal relies on an extra contact to
provide the second channel. It is recognized however that this would
introduce cross exchange terms between the three leads which do not allow
for equivalent coupling of the two channels. In fact, any diagonalization of
the scattering problem of the kind of the one outlined in the previous
Subsection to isolate the independent channels can never produce equal
coupling as long as off-diagonal terms are non zero. To overcome this
difficulty the authors propose that the extra contact is a larger dot
itself, having a charging energy $E_{c}$ which hinders exchange coupling
with the other two leads. Of course its size should be appropriately tuned.
A smaller size implies a larger level spacing $\delta $ inside it and $%
\delta $ should be low enough because it acts as the low energy cutoff in
the scaling toward the fixed point. Therefore, there is a delicate trade-off
between incresing $E_{c}$ to prevent cross exchange terms and decreasing $%
\delta $ not to stop the flow when reducing the temperature.

We have stressed that the NFL fixed point in the two channel Kondo coupling
at $T=0$ can only be reached by tuning the exchange couplings of both
channels $J_{1},J_{2}$ exactly equal. Hopefully this can be done by changing
an appropriate gate voltage across a critical point $V_{g}^{\ast }$.
According to Fig. 4 the systems flows to $J_{1}\rightarrow \infty
,J_{2}\rightarrow 0$ for $V_{g}<V_{g^{\ast }}$ and to $J_{2}\rightarrow
\infty ,J_{1}\rightarrow 0$ for $V_{g}>V_{g^{\ast }}$, both of which are
Fermi liquid fixed points. It has been argued that the quantum transition
across $V_{g}^{\ast }$ should display a quantum critical region in the $%
T-V_{g}$ plane whose critical properties can be determined \cite{glazman3}.
This offers better chances to spot whether we are in the vicinity of the NFL
fixed point or not, also at finite temperature.

\section{Summary}

In this review we focused on equilibrium transport properties across a
quantum dot (QD). A QD is a tunneling center for electrons coming from the
leads. Depending on the transparency of the barriers and on the temperature,
the coupling $V$ to the leads is weak or strong.

In the weak coupling regime, tunneling can be dealt with perturbatively. In
the case of the QD, a Coulomb Blockade zone is delimited by two sharp
conduction peaks which grow as $|V|^{2}/T$ in lowering the temperature.
Conduction inbetween is due to cotunneling processes and is exponentially
damped in temperature. Differential conductance is quite small, being $%
\propto |V|^{4}/U$, so that the charge degree of freedom is freezed on the
dot.

In lowering the temperature, non perturbative coupling of the dot to the
delocalized electrons of the contacts can occur and conductance can reach
the unitary limit in a CB valley. For pedagogical reasons we have reviewed
the old-fashoned Anderson-Yuval model for the correlated state. Otherwise we
have used the {\sl poor man's scaling }approach in the regime where
perturbative scaling applies and the Nozi\'{e}res scattering approach which
entails the fixed point physics at zero temperature. These approaches do not
allow for quantitative results which are better obtained with the numerical
renormalization group (NRG), real time RG and Non Crossing Approximation
(NCA) (the last one preferably in the overscreened case when a non Fermi
liquid (NFL) fixed point is reached), but they offer a more transparent view
of what is going on.

We have briefly reviewed the various types of Kondo couplings.

We have considered the case in which the GS of the dot is degenerate because
of spin: if temperature is low enough ($T<T_{K} $, with $T_{K}$ depending on
the transparency of the barriers), spin flip processes proliferate and the
magnetic moment of the dot is partially or fully screened . In this case an
applied magnetic field has a disruptive effect on the Kondo peak of the
conductance. The interesting case of a crossing between different dot spin
states induced by the magnetic field has been also discussed.

We have also reported on other possible realizations of Kondo physics
involving orbital degrees of freedom (orbital Kondo). The most favourable
set up for this case is a vertical geometry of the dot and the contacts with
cylindrical symmetry. In this geometry a magnetic field orthogonal to the
dot (which may be strong) can induce level crossings and produce the
degeneracies of the dot GS which is required for Kondo conductance to take
place. Hence one can have Kondo effect also with an even number of electrons
on the dot and zero total spin.

Some attention has been devoted to the multichannel Kondo effect. The
overscreening case can lead to a NFL fixed point at zero temperature. In
particular we have discussed the two channel spin $1/2$ Kondo state and
reported on possible experimental realizations that have been proposed. It
emerges that conditions to be met are very tough. Nonetheless its
achievement would probe a beautiful piece of the physics of the strongly
correlated systems.

\section{Appendix A: RG equations for the Coulomb gas model of eq. (29)}

We summarize here the scaling of the action in eq. (\ref{part1}) in order to
find out the behaviour of the system at large time scales (low temperature)
\cite{yuval}\cite{ni}. The cutoff is rescaled according to: $\tau
_{M}\rightarrow (1+\lambda )\tau _{M}$ with $\lambda =\Delta \tau _{M}/\tau
_{M}\rightarrow d\ln \tau _{M}$. This adds a factor $e^{-2N\lambda (1-\alpha
^{2}/2)}$. The first term arises from $\tau _{M}^{2N}$ in the denominator,
while the one $\propto \alpha ^{2}$ arises from the $\ln -$ term in the
interaction term. Indeed charge neutrality implies that $0=(%
\sum_{i}q_{i})^{2}=\sum_{i\neq j}q_{i}q_{j}+\sum_{i}q_{i}^{2}$ and $%
q_{i,j}=\pm 1$. This factor renormalizes the fugacity $Y\rightarrow Y+dY$,
with:
\begin{equation}
Y+dY\approx Ye^{\lambda (1-\alpha ^{2}/2)},  \label{yren}
\end{equation}
what gives the first of eqs. (\ref{rg1}). The interaction strength $\alpha $
is renormalized by flip-antiflip (particle-antiparticle) fusion.

Let us consider now all the configurations in which pairs of neighboring
charges $q_{i}$ at $\tau _{i}$ and $q_{j}=-q_{i}$ at $\tau _{j}$ (where $%
j=i\pm 1$) are at a distance between $\tau _{M}$ and $\tau _{M}(1+\lambda )$%
. In increasing the scale, these pairs are seen as a neutral compound which
screens the interaction between other charges (``fusion ''\ of pairs). Let
us consider one single fusion process and develop that part of the action
that contains their coordinates:
\begin{eqnarray}
-S^{(2N+2)} &=&-S^{(2N)}+q_{i}q_{j}\ln \left| \frac{\epsilon }{\tau _{M}}%
\right|   \nonumber \\
&&+\frac{q_{i}}{2}\sum_{k\neq i,j}q_{k}\left[ \ln \left| \frac{\tau
_{k}-\tau -\epsilon /2}{\tau _{M}}\right| -\ln \left| \frac{\tau _{k}-\tau
+\epsilon /2}{\tau _{M}}\right| \right] .
\end{eqnarray}
Here we have defined $\epsilon /2=\frac{\tau _{i}-\tau _{j}}{2}$ and $\tau =%
\frac{\tau _{i}+\tau _{j}}{2}$. The integral over $\tau $ and $\epsilon $
for $\epsilon /\tau _{M}<<1$ which appears in the partition function of eq. (%
\ref{part1}) is:
\begin{eqnarray}
&&e^{-S^{(2N)}}Y^{2}\int_{\tau _{i-1}+\tau _{M}}^{\tau _{j+1}-\tau _{M}}%
\frac{d\tau }{\tau _{M}}\int_{\tau _{M}}^{\tau _{M}(1+\lambda )}\frac{%
d\epsilon }{\tau _{M}}\left( \frac{\epsilon }{\tau _{M}}\right) ^{-\alpha
^{2}}{\cdot} e^{-\alpha ^{2}\epsilon \frac{q_{i}}{2}\sum_{k\neq i,j}q_{k}\frac{%
\partial }{\partial \tau }\ln \left| \frac{\tau _{k}-\tau )}{\tau _{M}}%
\right| }  \nonumber \\
&\sim &\lambda Y^{2}\left\{ 1-\alpha ^{2}\left( \frac{q_{i-1}}{2}\sum_{k\neq
i,j}^{2N}q_{k}\ln \left| \frac{\tau _{i-1}+\tau _{M}-\tau _{k})}{\tau _{M}}%
\right| +\frac{q_{j+1}}{2}\sum_{k\neq i,j}^{2N}q_{k}\ln \left| \frac{\tau
_{j+1}-\tau _{M}-\tau _{k})}{\tau _{M}}\right| \right) \right\} ,
\end{eqnarray}
where the term with $k=i-1$ $(k=j+1)$ in the first (second) sum vanishes.
Doing the same for each interval $i-j$ and summing, each term is repeated
twice. This generates an extra contribution to the action that can be
interpreted as the renormalization of the coupling constant:
\begin{equation}
\alpha ^{2}\rightarrow \alpha ^{2}+d\alpha ^{2}\;\;;\;d\alpha ^{2}=-2\lambda
Y^{2}\alpha ^{2}.  \label{aren}
\end{equation}
Because $\lambda \approx d\ln \tau _{M}$, the second one of eqs. (\ref{rg1})
is obtained.

\section{ Appendix B: Temperature dependence of the conductance}

Ohm's law can be derived semiclassically from a Boltzmann equation which
accounts for weak scattering of the charge carriers against impurities and
defects, when driven by an electric field. Scattering provides a mechanism
for relaxation from a non equilibrium to a steady state flow \cite{ashcroft}%
. We follow here a simplified approach starting from the velocity of band
electrons $\vec{v}_{k}=\vec{\nabla}\epsilon _{k}/\hbar $. The current
density is:
\begin{equation}
\vec{j}=-\frac{2e}{3{\cal {V}}}\sum_{k}\vec{v}_{k}(f_{k}-f_{k}^{o})
\end{equation}
where $f_{k}(f_{k}^{o})$ is the non-equilibrium (equilibrium) Fermi
distribution and ${\cal {V}}$ is the volume. Now we take:
\begin{equation}
f_{k}-f_{k}^{o}=-\frac{\partial f^{o}}{\partial \epsilon _{k}}\delta
\epsilon _{k}\hspace*{0.5cm}\delta \epsilon _{k}=-e\vec{E}{{\cdot} }\vec{v}%
_{k}\tau (k),
\end{equation}
which defines the relaxation time $\tau (k)$, so that, according to Ohm's
law, we get:
\begin{equation}
\sigma =-\frac{2e^{2}}{3{\cal {V}}}\sum_{k}\vec{v}_{k}{{\cdot} }\vec{v}
_{k}\tau (k)\frac{\partial f^{o}}{\partial \epsilon _{k}}.  \label{singult}
\end{equation}

In th case of $s-$ wave scattering a spherical wave is diffracted from the
impurity with maximum amplitude; this increases the flux propagating
backward. Assuming that quantities depend on $\epsilon $ only and are mostly
evaluated at the Fermi level we have ($T\sim 0$):
\begin{equation}
\sigma =-\frac{2e^{2}}{3{\cal {V}}}\nu (0){v}_{F}^{2}\int d\epsilon \: \tau
(\epsilon ,T)\frac{\partial f^{o}}{\partial \epsilon }.  \label{rising}
\end{equation}
In this way we recover the simple microscopic formula for Ohm's
conductivity, first derived by Drude in 1900, $\sigma =ne^{2}\tau /m$, where
$\tau $ is an average relaxation time. The latter can be defined in terms of
the mean free path between two scattering events $l=v_{F}\tau $ or distance
between impurities. This formula is valid in the limit $k_{F}l>>1$ which we
assume to be the case. In such a limit no localization effects occur and one
can show that it is $d\rho /dT>0$ always, which is violated in the case of
Kondo conductivity in diluted alloys.

In fact, for the large $U$ Anderson model, the explicit dependence of the
transport time on temperature at the Fermi energy has to be taken into
account separately. In addition, the Sommerfeld expansion can be used at
finite temperatures \cite{costi}:
\begin{equation}
\int d\epsilon \left( -\frac{\partial f(\epsilon )}{\partial \epsilon }%
\right) \tau (\epsilon ,T)=\tau (\mu ,T)+\left. \frac{\pi ^{2}}{6}%
(k_{B}T)^{2}\frac{d^{2}\tau }{d\epsilon ^{2}}\right| _{\epsilon =\mu }=\tau
(\mu ,0)\left[ 1+\frac{\pi ^{2}}{16T_{K}^{2}}\left( \frac{1}{2}\pi ^{2}T^{2}+%
\frac{\pi ^{2}}{6}3T^{2}\right) \right] .  \label{sommerfeld}
\end{equation}
The main ingredients to derive this formula are the Fermi liquid nature of
the zero temperature GS and the Taylor expansion of the temperature
dependence of the relaxation time. The prefactors have been specialized to
the case of the symmetric Anderson model (so that the lifetime of the Kondo
resonance is $\Gamma =4k_{B}T_{K}/\pi $).

It follows that, in diluted alloys, the resistivity takes a maximum at $T=0$%
:
\begin{equation}
\frac{\rho (T)}{\rho (0)}=1-\frac{\pi ^{4}T^{2}}{16T_{K}^{2}}.
\end{equation}
The extra correction on the r.h.s. of eq. (\ref{sommerfeld}) arises from
inelastic scattering at finite temperature, as shown by Nozi\'{e}res \cite
{nozieres}, according to the following argument.

In an elastic resonant scattering process at $T=0$ is $-\Im m\{t^{0}\}=\frac{%
1}{\nu (0)\pi }\sin ^{2}\delta ^{0}$. In a diffusive $s-$wave scattering at
finite $T$, we have to consider the total relaxation rate, defined as:
\[
\frac{1}{\tau (\epsilon )}=\sum_{k^{\prime }}w_{kk^{\prime }}(1-k{{\cdot} }%
k^{\prime })=\frac{2\pi }{\hbar }N_{i}\nu (0)\int d\epsilon _{k^{\prime
}}\delta (\epsilon _{k}-\epsilon _{k^{\prime }})|<k|t|k^{\prime }>|^{2}
\]
($s-$wave scattering implies that the correction $k{{\cdot} }k^{\prime }$
averages to zero). The inelasticity can be accounted for by defining an
effective elastic $S-$matrix which is now no longer unitary and an effective
phase shift which is different from the one at $T=0$.

The full $S-$matrix, instead, is of course unitary: $1=\sum_{\beta
}|S_{\alpha \beta }|^{2}$ $\equiv \sum_{\beta }[\delta _{\alpha \beta }-2\pi
i\nu (0)t_{\alpha \beta }\delta (\epsilon _{\alpha }-\epsilon _{\beta
})][\delta _{\alpha \beta }+2\pi i\nu (0)t_{\alpha \beta }^{\ast }\delta
(\epsilon _{\alpha }-\epsilon _{\beta })]$ $=\sum_{\beta }\delta _{\alpha
\beta }+\sum_{\beta }4\pi ^{2}\left( \nu (0)\right) ^{2}|t_{\alpha \beta
}|^{2}\delta (\epsilon _{\alpha }-\epsilon _{\beta }).$ Now, we separate the
elastic channels $\beta =\alpha $ from the inelastic ones on the r.h.s. and
we introduce the inelastic cross-section $W_{\alpha }^{in}=2\pi \sum_{\beta
\neq \alpha }\nu (0)|t_{\alpha \beta }|^{2}\delta (\epsilon _{\alpha
}-\epsilon _{\beta })$; hence, from unitarity it follows that $1+4\pi
^{2}\left( \nu (0)\right) ^{2}|t_{\alpha \alpha }|^{2}+2\pi \nu (0)W_{\alpha
}^{in}=1$. Thus, we can define an effective phase shift (which we call $%
\delta $ again) in presence of inelastic scattering by introducing an
effective \textquotedblleft elastic\textquotedblright\ $S$ matrix:
\begin{equation}
1-2\pi i\nu (0)t_{\alpha \alpha }=\left( 1-2\pi \nu (0)W_{\alpha
}^{in}\right) ^{\met}e^{2i\delta },
\end{equation}
where the square root can be expanded.

According to the optical theorem, we can write the total transmission
probability as:

\begin{equation}
W_{\alpha }=-2\Im m\{t_{\alpha \alpha }\}=W_{\alpha }^{in}\cos 2\delta +%
\frac{(1-\cos 2\delta )}{\nu (0)\pi }.  \label{pt1}
\end{equation}
Now, let us notice that, because $\delta \sim \pi /2$, we have $\cos 2\delta
<0$. Therefore, for each channel $\alpha $ the quantity $W_{\alpha }$ is
always smaller than the elastic case value: $2\sin ^{2}\delta /(\nu (0)\pi )$%
. But one can thermally average over the $\alpha $ channels; according to
the usual phase space argument which is invoked when scattering occurs close
to the Fermi surface at very low $T$, we obtain an average $\overline{W^{in}}
$ proportional to $T^{2}$: $\overline{W^{in}}\sim AT^{2}$. This gives rise
to the first term in the expansion of eq. (\ref{sommerfeld}). The second
term can be viewed as contributing to the Sommerfeld expansion with $\delta
(\epsilon )=\met\pi -a\epsilon $ from which it follows that $\sin ^{2}\delta
\sim 1-a^{2}\epsilon ^{2}$.

In the case of resonant tunneling across the dot, the conductance is given
by eq. (\ref{correc1}). Using the result of eq. (\ref{pt1}) directly one
obtains that conductance has a maximum at zero temperature and first
corrections in temperature are again of ${\cal {O}}(T^{2})$ (see eq. (\ref
{cstrong})).

\bigskip

{\bf Acknowledgements}

This manuscript is the revised and updated version of the lectures delivered
by one of us (A. T.) at INFN Laboratories in Frascati (Italy) during the
school ``Nanotubes \& Nanostructures'' (October 18-27, 2001). A. T. wishes
to thank S. Bellucci and M. De Crescenzi for the invitation in such a warm
and stimulating atmosphere. A. Naddeo was supported by a CNR fellowship
while this work was done.

\bigskip

{\bf Figure Captions}

\begin{itemize}
\item  Figure 1: Schematic drawing for a dot and the contacts in a vertical
setup; possibly a magnetic field is applied along the axis.

\item  Figure 2: Differential conductance on a gray scale as a function of
both $V_{g}$ and $V_{ds}$; the Kondo effect shows up near $V_{ds}=0$ \cite
{kastner}.

\item  Figure 3: Renormalization-group flow diagram for small $J$ \cite
{yuval}.

\item  Figure 4: Qualitative renormalization-group flow diagram for the
anisotropic two channel Kondo problem; the non trivial fixed point
corresponds to $J_{1}=J_{2}=J_{\ast }$ (channel symmetry) \cite{fabrizio}.

\item  Figure 5: Plot of the universal function $f\left( x\right) $ vs $%
x=T/T_{K}$ \cite{costi1}.

\item  Figure 6: Sketch of the conductance across the dot as a function of
the temperature $T$ and the gate voltage $V_{g}$ \cite{arturo3}.

\item  Figure 7: Second-order diagrams ${\cal D}_{1}$ and ${\cal D}_{2}$.
Conduction fermions are represented as full lines, while dashed lines
represent propagation of $d$-fermions (dot's states).

\item  Figure 8: Third-order vertex renormalization: the first one ($P$) is
a ''parquet-type'' diagram. Its contribution is accounted for in the
integration of the second-order RG equation. The second one ($NP$) is a
''non parquet'' diagram. It provides an additional third-order contribution
to the $\beta $-function.
\end{itemize}

\begin{figure}[tbp]
\centering \includegraphics*[width=0.5\linewidth]{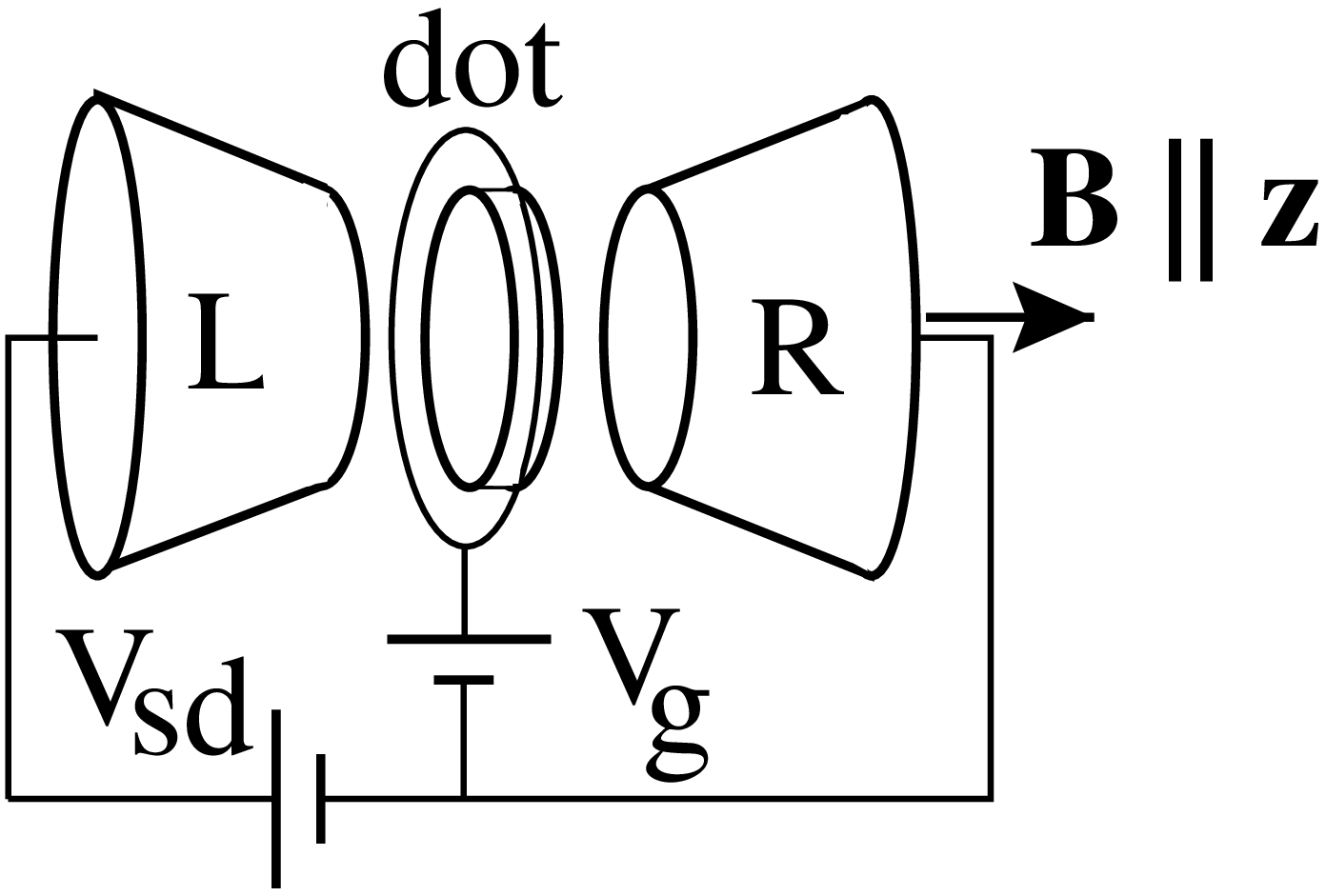}
\label{figura1}
\end{figure}
\begin{figure}[tbp]
\centering \includegraphics*[width=0.5\linewidth]{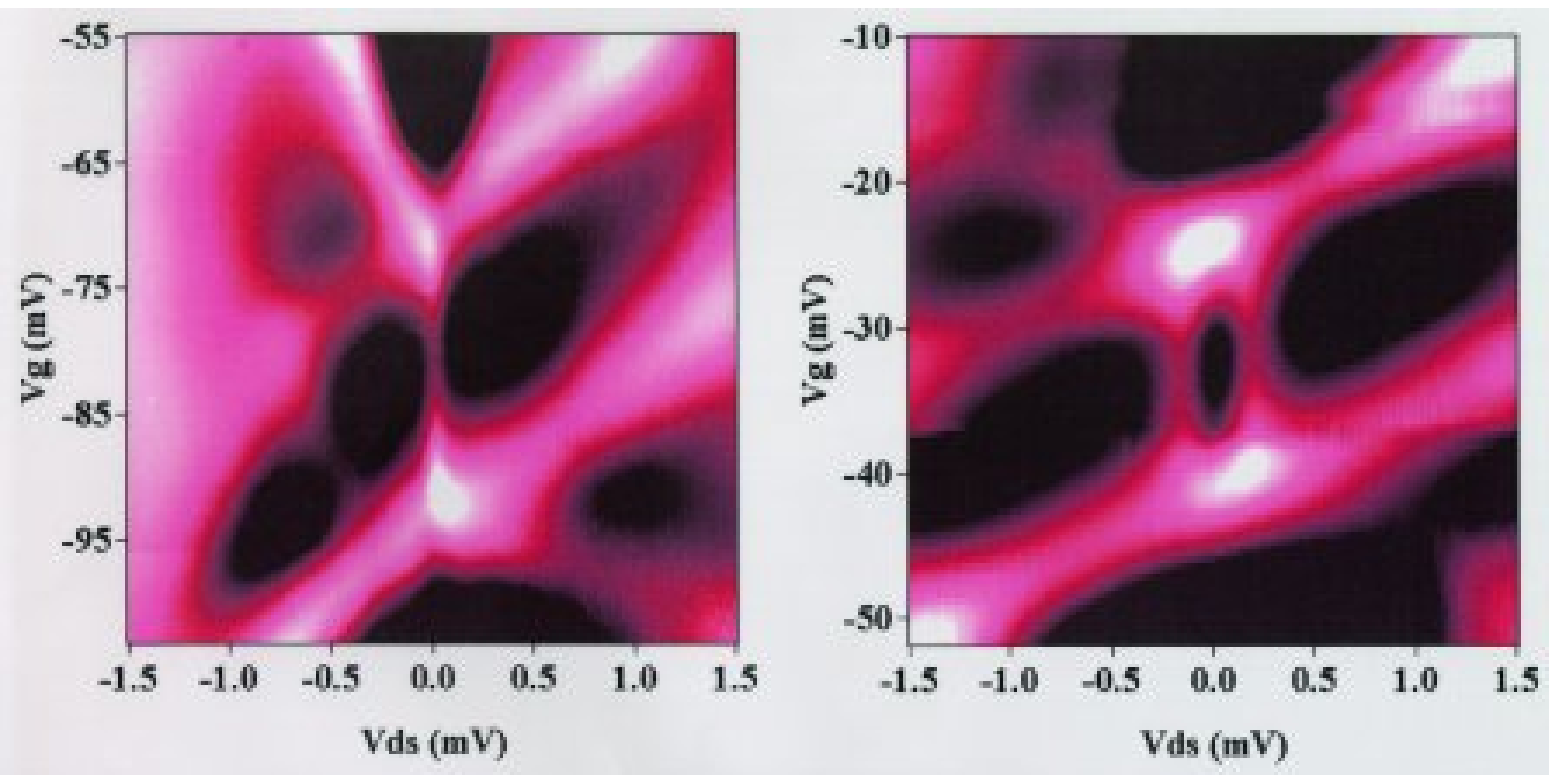}
\label{figura2}
\end{figure}
\begin{figure}[tbp]
\centering \includegraphics*[width=0.5\linewidth]{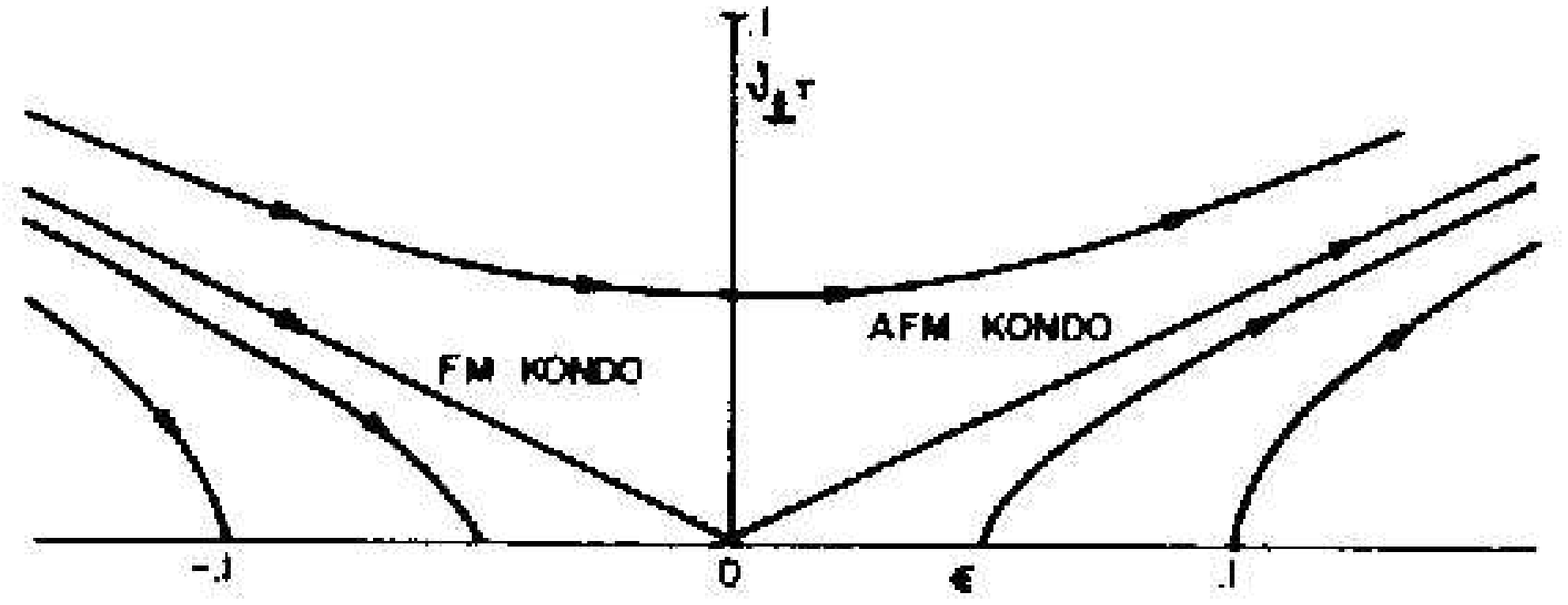}
\label{figura3}
\end{figure}
\begin{figure}[tbp]
\centering \includegraphics*[width=0.5\linewidth]{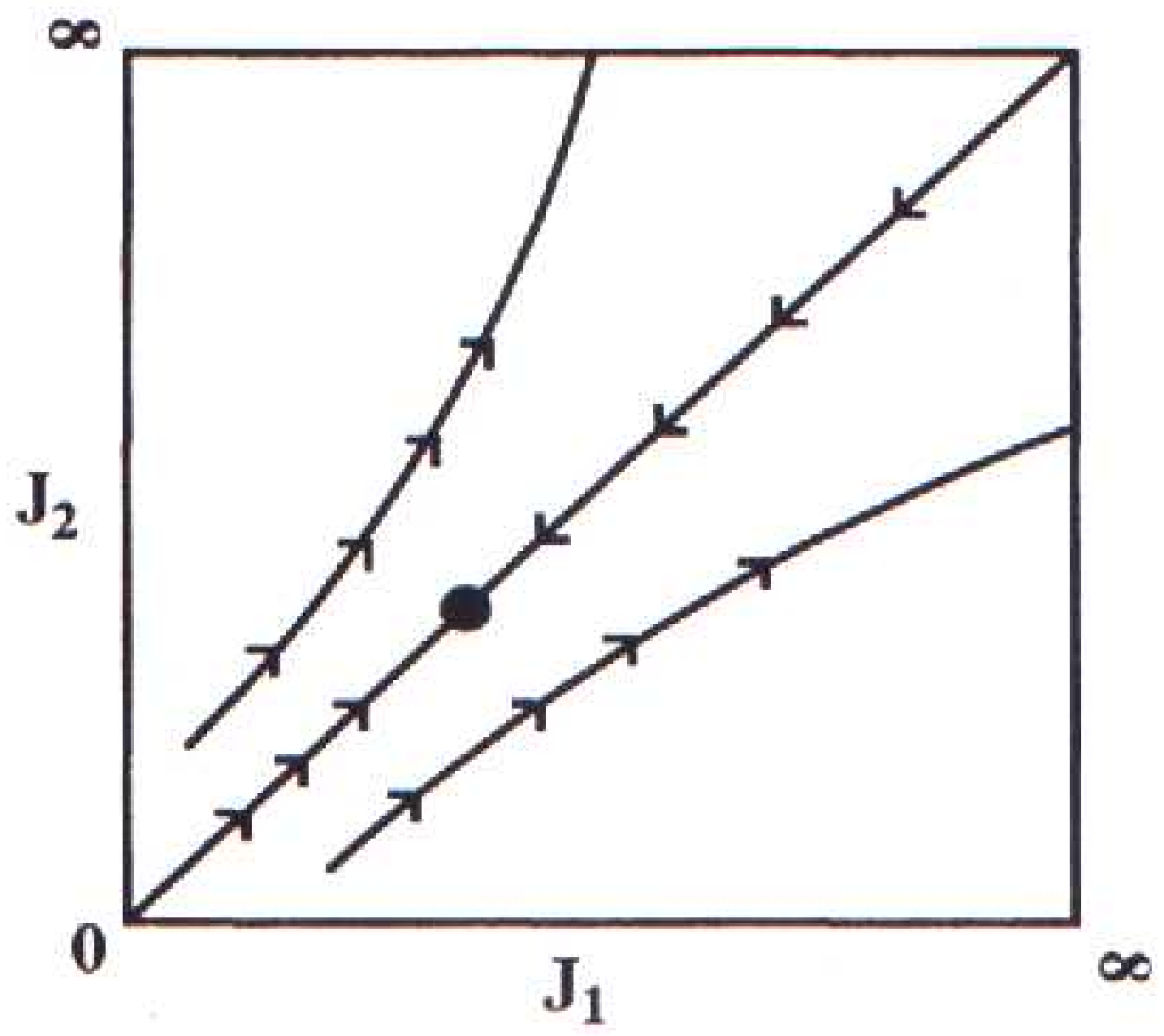}
\label{figura4}
\end{figure}
\begin{figure}[tbp]
\centering \includegraphics*[width=0.5\linewidth]{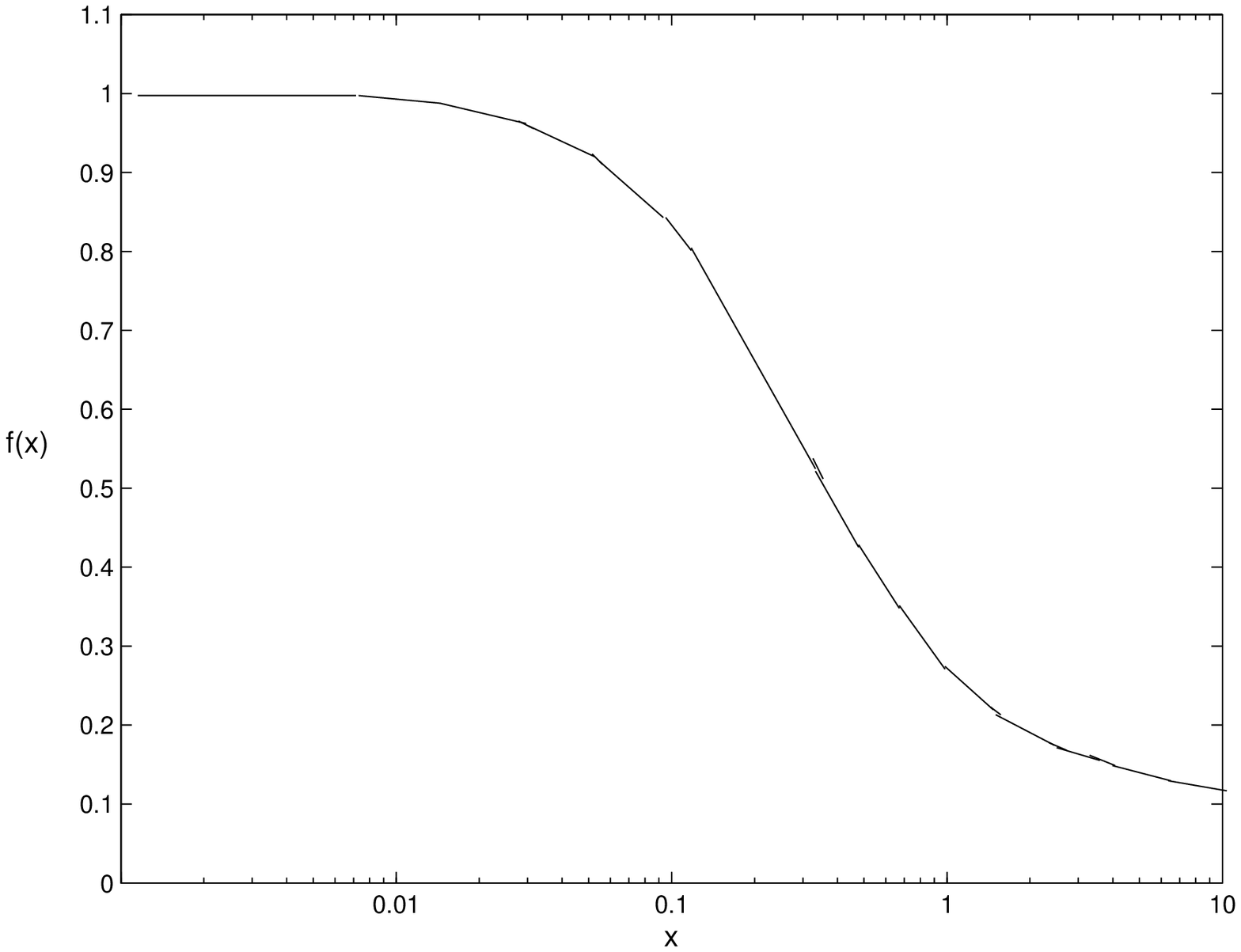}
\label{figura5}
\end{figure}
\begin{figure}[tbp]
\centering \includegraphics*[width=0.5\linewidth]{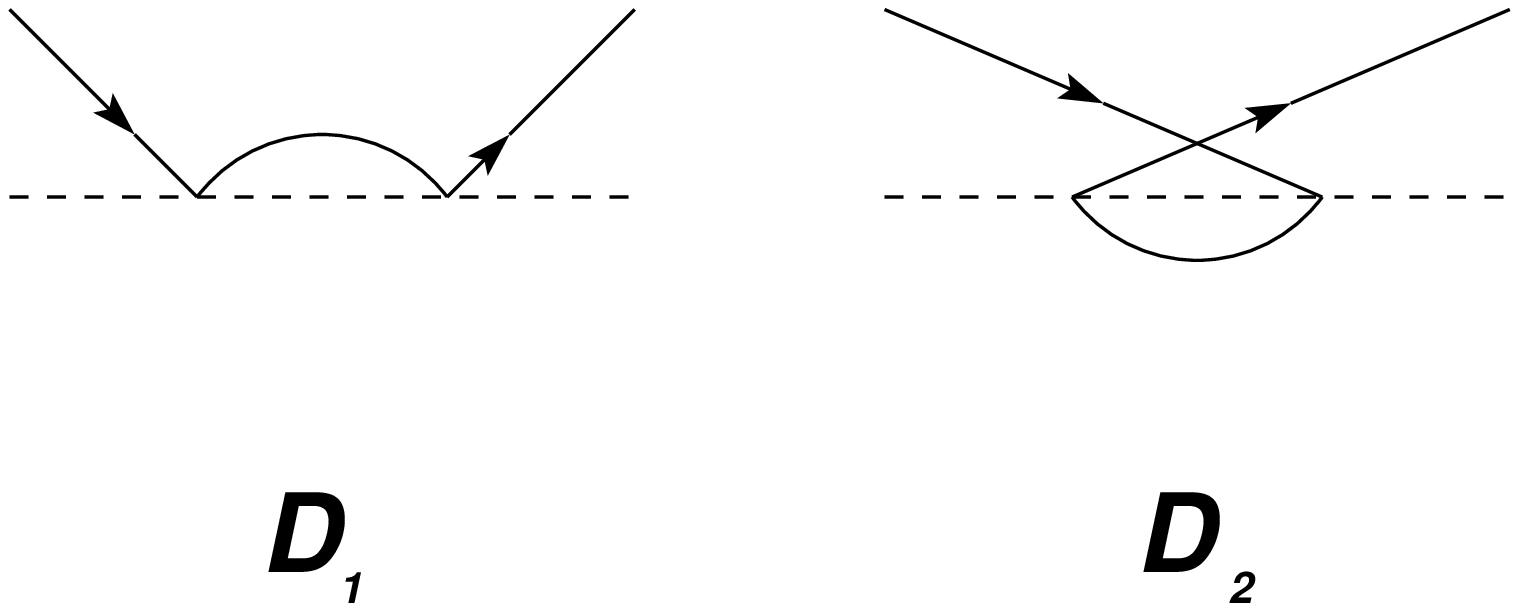}
\label{figura7}
\end{figure}
\begin{figure}[tbp]
\centering \includegraphics*[width=0.5\linewidth]{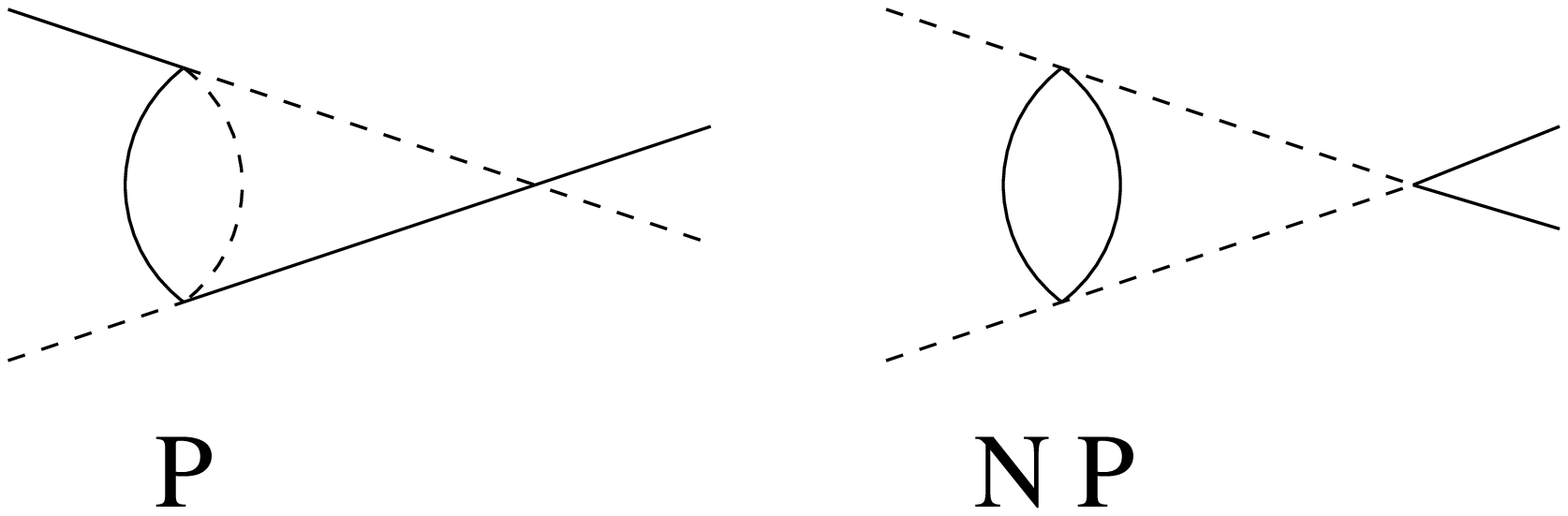}
\label{figura8}
\end{figure}

\end{document}